\documentclass{aa}
\usepackage{graphicx,amsfonts,amssymb,txfonts,natbib,latexsym,deluxetable}
\newcommand{\plotwd}{8.2cm}

\titlerunning{SDSS LRG acoustic oscillations} 
\title{Acoustic oscillations in the SDSS DR4 Luminous Red Galaxy sample power spectrum}
\author{G. H\"utsi\inst{1,2}}

\institute{Max-Planck-Institut f\"ur Astrophysik, Karl-Schwarzschild-Str. 1,
86740 Garching bei M\"unchen, Germany
\and Tartu Observatory, T\~oravere 61602, Estonia}

\offprints{G.H\"utsi , \\ \email{gert@mpa-garching.mpg.de}}
\date{Received / Accepted}
\begin{document}
\abstract {We calculate the redshift-space power spectrum of the Sloan Digital Sky Survey (SDSS) Data Release 4 (DR4) Luminous Red Galaxy (LRG) sample, finding evidence for a full series of acoustic features down to the scales of $\sim 0.2\,h\,\mathrm{Mpc}^{-1}$. This corresponds up to the 7th peak in the CMB angular power spectrum. The acoustic scale derived, $(105.4 \pm 2.3)\,h^{-1}\,\mathrm{Mpc}$, agrees very well with the ``concordance'' model prediction and also with the one determined via the analysis of the spatial two-point correlation function by \citet{2005ApJ...633..560E}. The models with baryonic features are favored by $3.3 \sigma$ over their ``smoothed-out'' counterparts without any oscillatory behavior. This is not only an independent confirmation  of \citet{2005ApJ...633..560E} results made with different methods and software but also, according to our knowledge, the first determination of the power spectrum of the SDSS LRG sample.
}
\maketitle

\section{Introduction}

\begin{figure}
\centering
\includegraphics[width=\plotwd]
{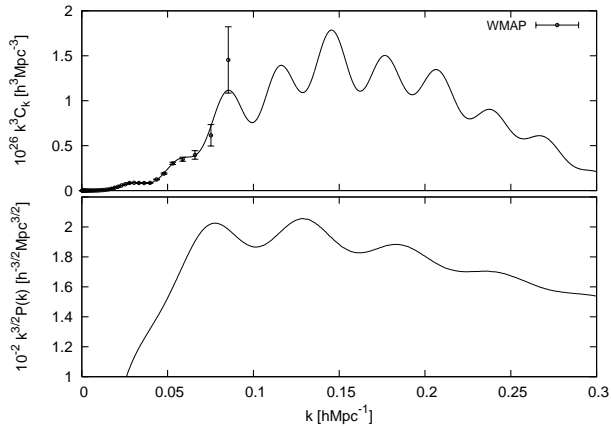}
\caption{Acoustic oscillations in the CMB (upper panel) and linear matter power spectrum (lower panel) for the ``concordance'' cosmological model. Here, as we have plotted the spectra against spatial wavenumber $k$, we have changed the standard notation of $C_\ell$ to $C_k$. Due to the $k^3$ factor the first CMB acoustic peak is barely visible. Density fluctuations in matter at smaller scales, being mostly induced by the velocity fields, are out of phase with respect to the fluctuations in the CMB component. Also the fluctuation period is twice as large.}
\label{fig1}
\end{figure}

In the beginning of $1970$'s it was already realized that acoustic waves in the tightly coupled baryon-photon fluid prior to the epoch of recombination will lead to the characteristic maxima and minima in the post-recombination matter power spectrum. The same mechanism is also responsible for the prominent peak structure in the CMB angular power spectrum \citep{1970Ap&SS...7....3S,1970ApJ...162..815P,1978SvA....22..523D}. The scale of these features reflects the size of the sound horizon, which itself is fully determined given the physical densities $\Omega_b h^2$ and $\Omega_m h^2$. The acoustic horizon can be calibrated using the CMB data, thus turning it into a standard ruler which can be used to carry out classical cosmological tests. For example, if we are able to measure the redshift and angular intervals corresponding to the physically known acoustic scale in the matter power spectrum at a range of redshifts, we can immediately find angular diameter distance $d_{\rm A}$ and Hubble parameter $H$ as a function of redshift. Having good knowledge of these dependencies allows us to put constraints on the properties of the dark energy. To carry out this project one needs a tracer population of objects whose clustering properties with respect to the underlying matter distribution is reasonably well understood. There have been several works discussing the usage of galaxies \citep{2003ApJ...594..665B,2003PhRvD..68f3004H,2003PhRvD..68h3504L,2003ApJ...598..720S} and clusters of galaxies \citep{2003PhRvD..68f3004H,2004ApJ...613...41M,astro-ph/0505441} for this purpose. What is most important is that already currently existing galaxy redshift surveys have lead to the detection of acoustic features in the spatial distribution of galaxies, this way providing clearest support for the standard gravitational instability picture of the cosmic structure formation. In the paper by \citet{2005ApJ...633..560E} the detection of the acoustic ``bump'' in the two-point redshift-space correlation function of the SDSS \footnote{http://www.sdss.org/} LRG sample is announced. The discovery of similar features in the power spectrum of 2dF \footnote{http://www.mso.anu.edu.au/2dFGRS/} galaxies is presented in \citet{2005MNRAS.362..505C}. These results clearly demonstrate the great promise of the future dedicated galaxy redshift surveys like K.A.O.S.\footnote{http://www.noao.edu/kaos/} Similarly, useful measurements of the acoustic scale can be hoped by the planned SZ cluster surveys like the ones carried out by the PLANCK Surveyor \footnote{http://astro.estec.esa.nl/Planck} spacecraft and SPT \footnote{http://astro.uchicago.edu/spt} \citep{astro-ph/0505441} and also with a large future photometric redshift surveys \citep{2005MNRAS.tmp..876B}. For the SZ surveys one needs an additional optical follow-up to get estimates for the cluster redshifts. In this paper we calculate the redshift-space power spectrum of the SDSS LRG sample finding evidence for the acoustic oscillations down to the scales of $\sim 0.2\,h\,\mathrm{Mpc}^{-1}$, which effectively correspond up to the 7. peak in the CMB angular power spectrum. These scales in the CMB are very strongly damped due to the finite width of the last-scattering surface and also due to the Silk damping \citep{1968ApJ...151..459S}. This can be seen in Fig. \ref{fig1} \footnote{Here instead of the usual multipole number $\ell$ we have plotted the CMB angular power spectrum against the wavenumber $k$. For the ``concordance'' cosmological model $\ell = 9990\,k[h\,\mathrm{Mpc}^{-1}]$.} where the CMB data is plotted in a somewhat unusual way to enhance the acoustic features at the high wavenumber damping tail. Also, at those scales the secondary CMB anisotropies (mostly thermal Sunyaev-Zeldovich effect \citep{1972CoASP...4..173S,1980ARA&A..18..537S}) start to dominate over the primary signal. On the other hand, features in the matter power spectrum, although being small ($\sim 5\%$ fluctuations), are preserved by the linear evolution and so opening up the way to probe acoustic phenomena at scales smaller than the ones accessible for the CMB studies. 

The paper is structured as follows. In Sec. 2 we describe the dataset to be analyzed. Sec. 3 presents the method of the power spectrum calculation. In Sec. 4 we determine power spectrum errors and covariance matrix. Sec. 5 discusses the convolution effect of the survey window. Analytical model spectra are presented in Sec. 6. The results of the measurement of the acoustic scale are given in Sec. 7. Correlation function analysis is carried out in Sec. 8. In Sec. 9 we compare the measured power spectrum with the published results for the 2dF and SDSS main sample, and finally we conclude with Sec. 10. 

\begin{figure}
\centering
\includegraphics[width=\plotwd]
{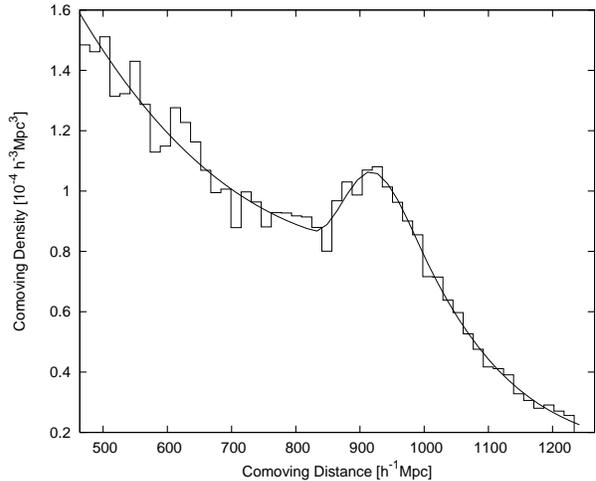}
\caption{Comoving number density of galaxies as a function of comoving distance. Smooth solid line shows a cubic spline fit to the number density estimated for 50 discrete radial bins.}
\label{fig2}
\end{figure}

\section{Data}
We analyze the publicly available data from the SDSS DR4 \citep{astro-ph/0507711}. Specifically, we carry out our power spectrum measurements using the subset of the SDSS spectroscopic sample known as the Luminous Red Galaxy (LRG) sample. The LRG selection algorithm \citep{2001AJ....122.2267E} selects $\sim 12$ galaxies per square degree meeting specific color and magnitude criteria \footnote{For the exact details of the selection criteria see \citet{2001AJ....122.2267E}}. The resulting set of galaxies consists mostly of an early types populating dense cluster environments and as such are significantly biased (bias factor $b \sim 2$) with respect to the underlying matter distribution. The selection method is very effective producing a galaxy sample with a reasonably high density up to the redshift of $z \sim 0.5$.

\begin{figure}
\centering
\includegraphics[width=\plotwd]
{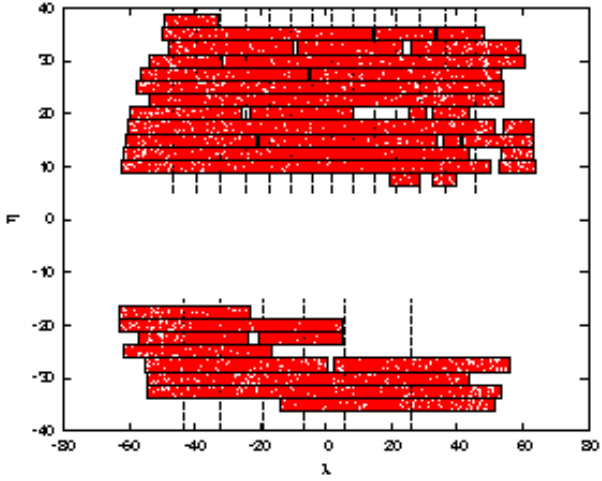}
\caption{Angular distribution of galaxies given in the SDSS survey coordinates $(\lambda,\eta)$. The survey mask is shown with solid lines. The vertical dashed lines show the division of the sample into 22 separate regions each containing $\sim 2350$ galaxies. This division will be exploited in the ``jackknife'' error analysis of the correlation function.}
\label{fig3}
\end{figure}

Since the selection criteria are very complicated, involving both cuts in magnitude and in color, and also due to the steepness of the luminosity function the usual method of using only the luminosity function to determine radial selection function does not work here \citep{2005ApJ...621...22Z}. Here we simply build the radial selection function as a smooth spline fit to the number density profiles shown in Fig. \ref{fig2}. To calculate distances we choose the cosmological parameters as given by the WMAP \footnote{http://lambda.gsfc.nasa.gov/product/map/} ``concordance'' model \citep{2003ApJS..148..175S}. Unfortunately the coverage masks of the SDSS DR4 spectroscopic sample are not available in a readily accessible format and so we chose to build the angular survey masks using the galaxy data itself \footnote{In principle one can build the angular masks using the raw tiling information, but as we show later our approximate treatment is probably rather fine, since the results seem to be quite stable against small uncertainties in the mask. More rigorous approach should certainly address the issues of survey boundaries and completeness fluctuations (expected to be small due to the very effective tiling algorithm by \citet{2003AJ....125.2276B}) in a much better detail.}. As the number density of galaxies in the sample is rather high, one can determine relatively accurately the beginning, ending and also possible gaps in the scan stripes. We have built angular masks using both the whole DR4 galaxy sample and LRGs only. The measured power spectra are practically identical with only some minor differences on smaller scales (see Fig. \ref{fig6}). This can be seen as an indication that our power spectrum measurements are rather stable against small uncertainties in the survey geometry. The angular distribution of the galaxies and also the boundaries of the survey mask built in the above mentioned way (here using all the galaxies) is shown in Fig. \ref{fig3}. Here the angular positions are plotted using the so-called survey coordinate system of the SDSS \footnote{The transformations between various coordinate systems used by the SDSS are given e.g. in \citet{2002AJ....123..485S}.}. 

We have selected all the objects that have spectrum classified as galaxy (i.e. \texttt{SpecClass=2}) and are additionally flagged as \texttt{GALAXY\_RED} or \texttt{GALAXY\_RED\_II} (i.e. \texttt{PrimTarget} bit mask set as \texttt{0x20} or \texttt{0x4000000}, respectively). Only galaxies for which the redshift confidence parameter, \texttt{zConf}, is greater than $0.95$ were used.
We apply lower and upper redshift cutoffs of $0.16$ and $0.47$ as also done in \citet{2005ApJ...633..560E}. The lower cutoff is needed since the color cuts that define the LRG sample break down for redshifts below $\sim 0.2$ \citep{2001AJ....122.2267E}. For the analysis presented in this paper we have excluded the three southern stripes since these just increase the sidelobes of the survey window without adding much of the extra volume. We have also removed some minor parts of the sample to obtain more continuous and smooth chunk of volume. In total the analyzed galaxy sample covers $\sim 0.75 \,h^{-3}\,\mathrm{Gpc}^3$ over $\sim 3850$ square degrees on the sky and contains $51,763$ galaxies. 

\section{Power spectrum calculation}
We calculate the power spectrum using a direct Fourier method as described in \citet{1994ApJ...426...23F} (FKP). Strictly speaking, power spectra determined this way are the so-called pseudospectra, meaning that the estimates derived are convolved with a survey window. Since in the case of the analyzed LRG sample the volume covered is very large, reaching $0.75 \,h^{-3}\,\mathrm{Gpc}^3$,  and also the survey volume has relatively large dimensions along all perpendicular directions, the correlations in the Fourier space are rather compact. On intermediate scales and in the case the power spectrum binning is chosen wide enough, FKP estimator gives a good approximation to the true underlying power.

The FKP estimate for a 3D pseudospectrum reads as:
\begin{equation}\label{eq2}
\tilde{P}(\mathbf{k}) = |F(\mathbf{k})|^2 - P_\mathrm{shot} \,, 
\end{equation}
where
\begin{equation}
F(\mathbf{k}) = \int {\rm d}^3r \, F(\mathbf{r})\exp(i\mathbf{k}\cdot\mathbf{r}) \,.
\end{equation}
Here $F(\mathbf{r})$ is the weighted density contrast field:
\begin{equation}\label{eq1}
F(\mathbf{r}) = w(\mathbf{r})\left[n_g(\mathbf{r})-\alpha n_s(\mathbf{r})\right] \,.
\end{equation}
$n_g(\mathbf{r})$ and $n_s(\mathbf{r})$ denote the number densities of the
analyzed galaxy catalog and a synthetic random catalog with the same selection
criteria, respectively. Since we are dealing with discrete point processes, densities can be given as:
\begin{eqnarray}
n_g(\mathbf{r}) = \sum \limits_i \delta^D(\mathbf{r}-\mathbf{r}^g_i)\,,\\
n_s(\mathbf{r}) = \sum \limits_i \delta^D(\mathbf{r}-\mathbf{r}^s_i)\,,
\end{eqnarray}
where $\mathbf{r}^g_i$ and $\mathbf{r}^s_i$ denote the location of the $i-$th point in real and synthetic catalog, respectively, and $\delta^D$ is the 3D Dirac delta function. $\alpha$ in Eq. (\ref{eq1}) is the ratio of the number of galaxies to the number of random points in the synthetic catalog i.e. $\alpha=\frac{N_g}{N_s}$. In our calculations we have $N_s=10^7$ and thus $\alpha \simeq 0.0052$. For the weight function $w(\mathbf{r})$ there have been traditionally three choices in the literature:
\begin{equation}
w(\mathbf{r})\propto \left\{ 
\begin{array}{lll}
\frac{1}{\bar{n}(\mathbf{r})} & \rm{\quad for\ volume\ weighting}\\
\rm{const} & \rm{\quad for\ number\ weighting}\\
\frac{1}{1+\bar{n}(\mathbf{r})\tilde{P}} & \rm{\quad for\ an\ optimal\ FKP\ weighting.}
\end{array} \right.
\end{equation}
Here $\bar{n}(\mathbf{r})$ is the average number density of galaxies at comoving location $\mathbf{r}$ i.e. the radial selection function of the survey (see Fig. \ref{fig2}) times the angular mask (Fig. \ref{fig3}). In our calculations we use an optimal FKP weighting scheme, although pure volume weighting would give practically the same results, especially on the larger scales ($k \lesssim 0.09\,h\,\mathrm{Mpc}^{-1}$), since then for the majority of the sample $\bar{n}(\mathbf{r})\tilde{P} \sim 3$ \footnote{Including all the modes down to the scales of $k \sim 0.25\,h\,\mathrm{Mpc}^{-1}$ the effective value for $\bar{n}(\mathbf{r})\tilde{P}$ drops down to $\sim 1.5$.}. The weights in Eq. (\ref{eq1}) are normalized such that:
\begin{equation}
\int {\rm d}^3r \,\bar{n}^2(\mathbf{r}) w^2(\mathbf{r}) = 1\,,
\end{equation}
which can be approximated as the following sum over the synthetic catalog \footnote{We assume that the survey selection does not have any other angular dependence except for the applied angular mask i.e. we can replace $\mathbf{r}^s_i$ by the modulus $r^s_i$.}:    
\begin{equation}
\alpha \sum \limits_i \bar{n}(r^s_i)w^2(r^s_i) = 1\,.
\end{equation}
The last term in Eq. (\ref{eq2}) represents the Poissonian discreteness noise and can be expressed as:
\begin{equation}
P_\mathrm{shot} = (1+\alpha) \int {\rm d}^3r \,\bar{n}(\mathbf{r}) w^2(\mathbf{r}) \simeq \alpha(1+\alpha)\sum \limits_i w^2(r^s_i)\,.
\end{equation}

Since we are using Fast Fourier Transforms (FFTs) to speed up the calculation of the Fourier sums, we have to deal with some extra complications. As the density field is now ``restricted to live'' on a regular grid with a finite cell size, we have to correct for the smoothing effect this has caused. Also, if our underlying density field contains spatial modes with higher frequency than our grid's Nyquist frequency, $k_{\mathrm{Ny}}$, then these will be ``folded back'' into the frequency interval the grid can support, increasing power close to $k_{\mathrm{Ny}}$-- the so-called aliasing effect. The relation between the spectra calculated using direct summation and the ones found using FFT techniques was worked out by \citet{2005ApJ...620..559J}. It can be expressed as follows:
\begin{eqnarray}\label{eq4}
|F(\mathbf{k})|_{\mathrm{FFT}}^2=\sum \limits_{\mathbf{n} \in \mathbb{Z}} |\mathcal{W}(\mathbf{k}+2k_{\mathrm{Ny}}\mathbf{n})|^2\tilde{P}(\mathbf{k}+2k_{\mathrm{Ny}}\mathbf{n}) + \nonumber \\
P_\mathrm{shot}\sum \limits_{\mathbf{n} \in \mathbb{Z}} |\mathcal{W}(\mathbf{k}+2k_{\mathrm{Ny}}\mathbf{n})|^2\,,
\end{eqnarray}
where $\mathcal{W}(\mathbf{k})$ is the mass assignment function used to build
density grid out of the point set. We use the Triangular Shaped Cloud (TSC)
assignment method \citep{1988csup.book.....H}. Since the TSC filter can be obtained by convolving uniform cube (the Nearest Grid Point filter) two times with itself, the Fourier representation of it follows immediately:
\begin{equation}
\mathcal{W}(\mathbf{k}) = \left[\frac{\prod \limits_{i=1}^3 \sin \left(\frac{\pi k_i}{2k_{\mathrm{Ny}}}\right)}{\left(\frac{\pi k_i}{2k_{\mathrm{Ny}}}\right)}\right]^3, \quad \mathbf{k} = (k_1,k_2,k_3)\,. 
\end{equation}
Here the sum that represents the contribution from aliases runs over all the integer vectors $\mathbf{n}$. Eq. (\ref{eq4}) is the direct analog of the previous Eq. (\ref{eq2}). The convolution with the mass assignment filter has introduced $\mathcal{W}^2(\mathbf{k})$ factors both to the power spectrum and to the shot noise term. The sum in the last term of Eq. (\ref{eq4}) can be performed analytically for the TSC filter to yield the result \citep{2005ApJ...620..559J} \footnote{For the NGP filter this sum equals 1, and so one recovers the original shot noise term in Eq.(\ref{eq2}).}:
\begin{equation}
\sum \limits_{\mathbf{n} \in \mathbb{Z}} |\mathcal{W}(\mathbf{k}+2k_{\mathrm{Ny}}\mathbf{n})|^2 = \prod \limits_{i=1}^3 \left[ 1-\sin^2\left(\frac{\pi k_i}{2k_{\mathrm{Ny}}}\right) + \frac{2}{15}\sin^4\left(\frac{\pi k_i}{2k_{\mathrm{Ny}}}\right)\right]\,.
\end{equation}
To recover the angle averaged pseudospectrum $\tilde{P}(k)$ from Eq. (\ref{eq4}) we use an iterative scheme as described in \citet{2005ApJ...620..559J} with a slight modification: we do not approximate the small scale spectrum by a simple power law, but also allow for the possible running of the spectral index i.e. the parametric shape of the power spectrum is taken to be a parabola in log-log. Since on small scales the power spectrum is dropping fast, the sum over $\mathbf{n}$ in Eq. (\ref{eq4}) is converging rather rapidly. In calculations we use only integer vectors with $|\mathbf{n}| \leq 5$. The angular average is taken over all the vectors $\mathbf{k}$ laying in the same $k$-space shell with width $\Delta k$. The resulting $\tilde{P}$ is taken to be an estimate for the pseudospectrum at the wavenumber $k_{\mathrm{eff}}$ that corresponds to the average length of the $k$-vectors in that shell. 

To summarize, our power spectrum calculation consists of the following steps:
\begin{enumerate}
\item Determination of the survey selection function i.e. mean underlying number density $\bar{n}(\mathbf{r})$ (including the survey geometry),
\item Calculation of the overdensity field on a grid using TSC mass assignment scheme,
\item Fourier transformation of the gridded density field,
\item Calculation of the raw 3D power spectrum $|F(\mathbf{k})|_{\mathrm{FFT}}^2$,
\item Subtraction of the shot noise component from the raw spectrum,
\item Recovery of the angle averaged pseudospectrum $\tilde{P}(k)$ using an iterative method of \citet{2005ApJ...620..559J}.
\end{enumerate}
 
We have applied the above described power spectrum calculation method to a multitude of test problems, the results of which can be found in \citet{astro-ph/0505441}. In Appendix \ref{appa} we show only one example, where we successfully recover the underlying power spectrum of galaxy clusters from the VIRGO Hubble Volume simulations \footnote{http://www.mpa-garching.mpg.de/Virgo/}, after applying the selection criteria given in Figs. \ref{fig2} and \ref{fig3}.

\section{Power spectrum errors and covariance matrix}
We determine power spectrum errors by three different methods:
\begin{enumerate}
\item Prescription given by FKP that assumes the underlying density field to be Gaussian. This method also does not treat redshift space distortions. Under those simplifying assumptions the power spectrum variance can be expressed as:  
\begin{eqnarray}
\sigma_{\tilde{P}}^2(k) = \frac{2}{N_k^2}\sum \limits_{\mathbf{k'}}\sum \limits_{\mathbf{k''}}|\tilde{P}(k)Q(\mathbf{k'}-\mathbf{k''})+S(\mathbf{k'}-\mathbf{k''})|^2\,,\label{eq5}\\
Q(\mathbf{k}) = \int {\rm d}^3r\,\bar{n}^2(\mathbf{r})w^2(\mathbf{r})\exp(i\mathbf{k}\cdot\mathbf{r}) \simeq \nonumber \\
\alpha \sum \limits_j \bar{n}(r^s_j)w^2(r^s_j)\exp(i\mathbf{k}\cdot\mathbf{r}^s_j)\,,\\
S(\mathbf{k}) = (1+\alpha)\int {\rm d}^3r\,\bar{n}(\mathbf{r})w^2(\mathbf{r})\exp(i\mathbf{k}\cdot\mathbf{r}) \simeq \nonumber \\
\alpha(1+\alpha)\sum \limits_j w^2(r^s_j)\exp(i\mathbf{k}\cdot\mathbf{r}^s_j)\,.
\end{eqnarray}
Here the sum is over all the wavevectors $\mathbf{k'}$ and $\mathbf{k''}$ populating the same $k$-space shell with radius $k$ and thickness $\Delta k$, and $N_k$ denotes the total number of modes in that shell. Since the direct summation over all the vector pairs $\mathbf{k'}$ and $\mathbf{k''}$ is very slow for the wide $k$-space shells and $512^3$ grid we use, a Monte Carlo sum is performed instead. Thus we calculate the average of the quantity $|\tilde{P}(k)Q(\mathbf{k'}-\mathbf{k''})+S(\mathbf{k'}-\mathbf{k''})|^2$ over the random pairs of vectors $\mathbf{k'}$ and $\mathbf{k''}$ from the same shell. For the result to converge properly we need on average $\sim 10^7$ random pairs.
\item The second method is a simple analytical approximation to the first one, also due to FKP (see also \citealt{1998ApJ...499..555T}). Here the variance is given as:
\begin{equation}
\sigma_{\tilde{P}}^2(k) = \frac{2\tilde{P}^2(k)}{V_{\rm eff}V_{\rm k}}\,, 
\end{equation}
where $V_{\rm k}=4\pi k^2\Delta k/(2\pi)^3$ is the volume of the $k$-space shell and $V_{\rm eff}$ is the effective volume given by:
\begin{equation}
V_{\rm eff} = \frac{\left[\int {\rm d}^3r \,\bar{n}^2(\mathbf{r}) w^2(\mathbf{r})\right]^2}{\int {\rm d}^3r \, \bar{n}^4(\mathbf{r}) w^4(\mathbf{r}) \left[1+ \frac{1}{\bar{n}(\mathbf{r})\tilde{P}(k)}\right]^2}\,.
\end{equation}
\item The third method is a Monte Carlo approach that uses $1000$ mock catalogs generated in the way described in Appendix \ref{appb}. Here, as we use the 2nd order Lagrangian perturbation theory, we get a good approximation for the mode-mode couplings that are induced during the quasi-nonlinear regime of the evolution of the density fluctuations. Also the large-scale redshift distortions are properly accounted for. In terms of the Halo Model (see Appendix \ref{appc}) we can say that halo-halo clustering term is relatively well approximated. Contributions from the one-halo term can be added later, as these allow an analytic treatment.     
\end{enumerate} 

\begin{figure}
\centering
\includegraphics[width=\plotwd]
{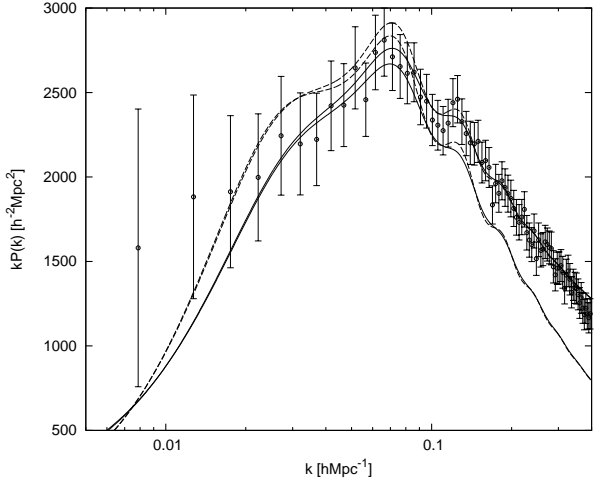}
\caption{Power spectrum of the SDSS LRG sample with the bin width $\Delta k \simeq 0.005 \,h\,\mathrm{Mpc}^{-1}$. The upper solid line shows the best fitting model spectrum and the lower one corresponds to the linearly evolved matter power spectrum of the ``concordance'' cosmological model multiplied by the square of the bias parameter $b=1.95$. Both of the spectra are convolved with a survey window. The dashed lines represent the corresponding unconvolved spectra.}
\label{fig4}
\end{figure}

\begin{figure}
\centering
\includegraphics[width=\plotwd]
{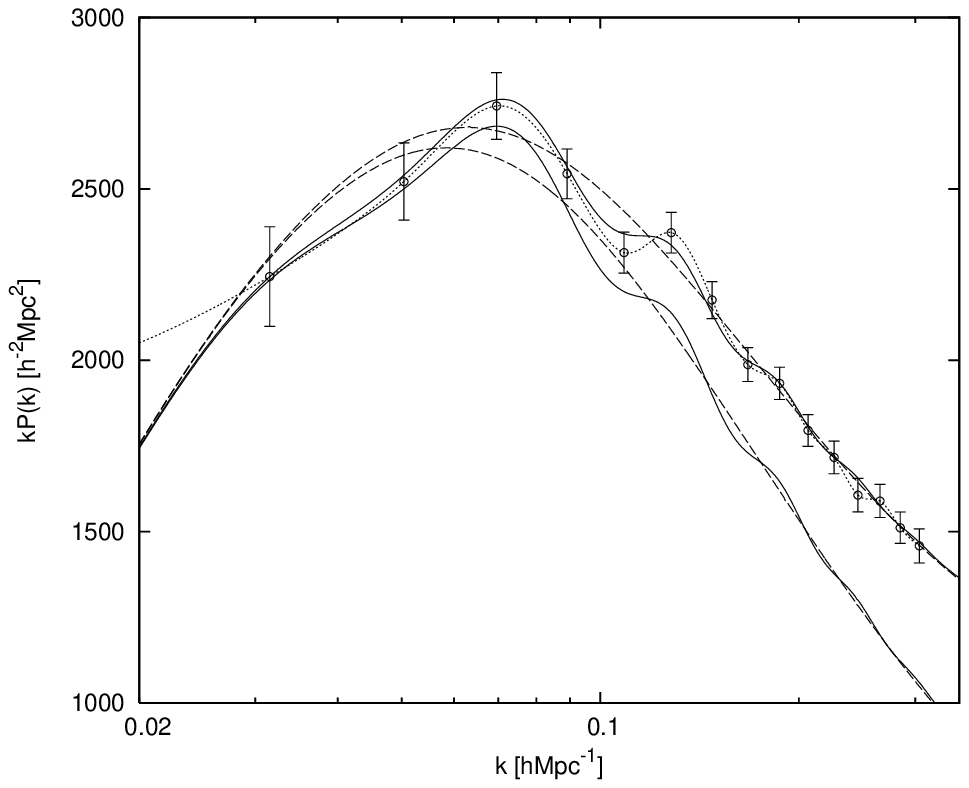}
\caption{Power spectrum of the SDSS LRG sample with the bin width $\Delta k \simeq 0.02 \,h\,\mathrm{Mpc}^{-1}$. The upper solid line shows the best fitting model spectrum and the lower one corresponds to the linearly evolved matter power spectrum of the ``concordance'' cosmological model multiplied by the square of the bias parameter $b=1.95$. Both of the spectra are convolved with a survey window. The dashed lines represent the ``smoothed-out'' versions of the above model spectra. The dotted line is the cubic spline fit to the data points.}
\label{fig5}
\end{figure}

\begin{figure}
\centering
\includegraphics[width=\plotwd]
{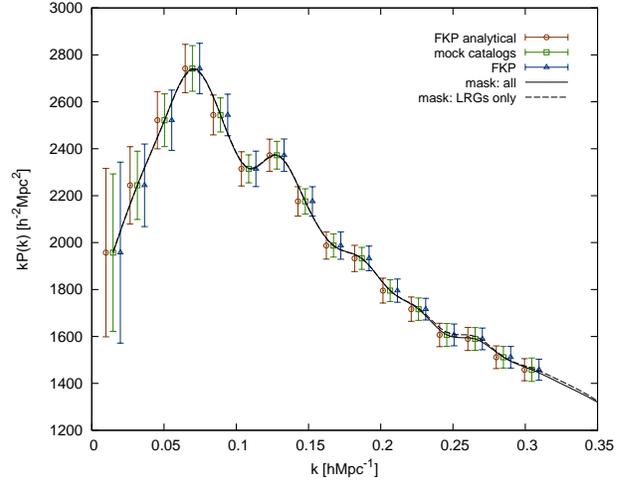}
\caption{The comparison of the different power spectrum error estimates. For clarity slight relative shifts of the data points have been applied. The errorbars resulting from the 1st method are the rightmost ones and the ones from the 3rd method are displayed in the middle. The lines show a cubic spline fits to the data points. The solid line corresponds to the case when all the available galaxy data is used to find the angular mask of the survey, while the dashed line represents the case when LRGs only are used for this purpose.}
\label{fig6}
\end{figure}

The results of the power spectra for the SDSS DR4 LRG sample are shown in Figs. \ref{fig4} and \ref{fig5}. In Fig. \ref{fig4} the bin width $\Delta k \simeq 0.005 \,h\,\mathrm{Mpc}^{-1}$, while in Fig. \ref{fig5} $\Delta k \simeq 0.02 \,h\,\mathrm{Mpc}^{-1}$. With different lines we have shown various model spectra, which will be the topic of Sec. \ref{sec6}.

The comparison of the power spectrum errorbars calculated in the above described different ways is provided in Fig. \ref{fig6}. We see that the various error estimates are in a very good agreement. In the following we will use only the errorbars given by the 3rd method.

So far we have only found the diagonal terms of the covariance matrix. In order to answer the question of how strongly different power spectrum bins are correlated, we have to go a step further, and try to estimate the full covariance matrix.

\begin{figure*}
\centering
\includegraphics[width=18cm]
{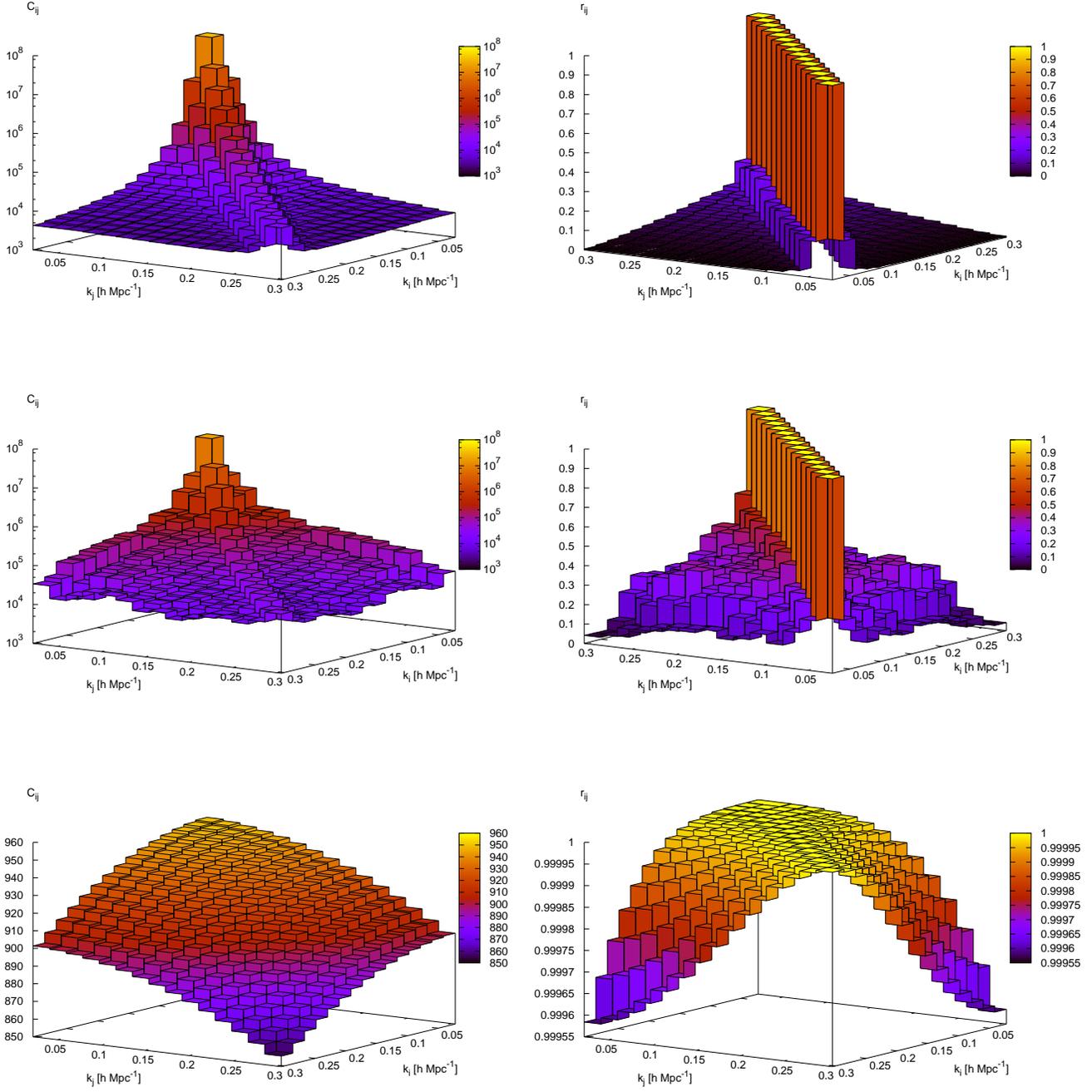}
\caption{Covariance (left column) and correlation matrices (right column). Top row represents the results from FKP prescription (see Eq. (\ref{eq6})) and the middle row the ones from $1000$ mock catalogs. The last row displays the nonlinear contribution due to the 1-halo term.}
\label{fig7}
\end{figure*}

The FKP result for the full covariance matrix, $C_{ij}$, is a simple generalization of the Eq. (\ref{eq5}):
\begin{equation}\label{eq6}
C_{ij}=\frac{2}{N_{k'}N_{k''}}\sum \limits_{\mathbf{k'}}\sum \limits_{\mathbf{k''}}\left|\tilde{P}\left(\frac{k_i+k_j}{2}\right)Q(\mathbf{k'}-\mathbf{k''})+S(\mathbf{k'}-\mathbf{k''})\right|^2\,,
\end{equation}
where the $k$-vectors $\mathbf{k'}$ and $\mathbf{k''}$ lie in shells with width $\Delta k$ and radii $k_i$ and $k_j$, respectively. The FKP approach, as mentioned above, does not treat mode couplings arising from the nonlinear evolution and also from the redshift space distortions. Linear redshift space distortions can be, in principle, included into the FKP estimate for the covariance matrix. One can generalize the results presented in the Appendix of \citet{1996ApJ...462...25Z}, where the covariance matrix for the Fourier modes has been found. Since linear redshift distortions applied on a Gaussian field do not change the Gaussianity property, one can still use the result from the Appendix B of FKP that relates the power spectrum covariance matrix to the covariance matrix of the Fourier modes. Also one has to add the shot noise terms to the result of \citet{1996ApJ...462...25Z}. We have carried out this exercise, leading us to the high dimensional integrals (up to 12 dim.) that turn out to be too time consuming to solve in practice. As from the mock catalogs we can hopefully obtain more realistic estimate for the covariance matrix \footnote{Since now we are also able to handle quasi-nonlinear mode-mode couplings.} we have not followed this path any further.

The results for the covariance matrix calculation are given in Fig. \ref{fig7}. Here the left hand column shows the covariance and the right hand column respective correlation matrices:
\begin{equation}
r_{ij}=\frac{C_{ij}}{\sqrt{C_{ii}C_{jj}}}\,.
\end{equation}
The power spectrum binning is the same as shown in Fig. \ref{fig5} i.e. $\Delta k \simeq 0.02 \,h\,\mathrm{Mpc}^{-1}$. The top row represents the results from Eq. (\ref{eq6}), while the middle row the ones from mock catalogs. Although the diagonal terms of the covariance matrices in the 1st and 2nd row are in a very good agreement (see Fig. \ref{fig6}), the off-diagonal components differ strongly. This can be explained as the result of the extra mode-mode couplings that are not accounted for by the FKP approach. We see that even well separated power spectrum bins can be correlated at $30 \ldots 40\%$ level. The bottommost row in Fig. \ref{fig7} represents the nonlinear contribution to the covariance matrix arising from the 1-halo term (see Appendix \ref{appc}). We see that this contribution is subdominant at the scales of interest to us \footnote{In calculating this contribution to the covariance matrix we have taken the best fit model parameters as obtained in Sec. \ref{sec6}. The smallness of this term is caused by the high value of the parameter $M_0$ i.e. majority of the ``occupied'' halos contain only one LRG.}. 

In the following calculations we mostly use the covariance matrix given in the middle row of Fig. \ref{fig7}.\footnote{This matrix along with the power spectrum results in Fig. \ref{fig5} is also given in a tabular form in Appendix \ref{appg}.}

\section{Relation to the true spectrum}
\begin{figure}
\centering
\includegraphics[width=\plotwd]
{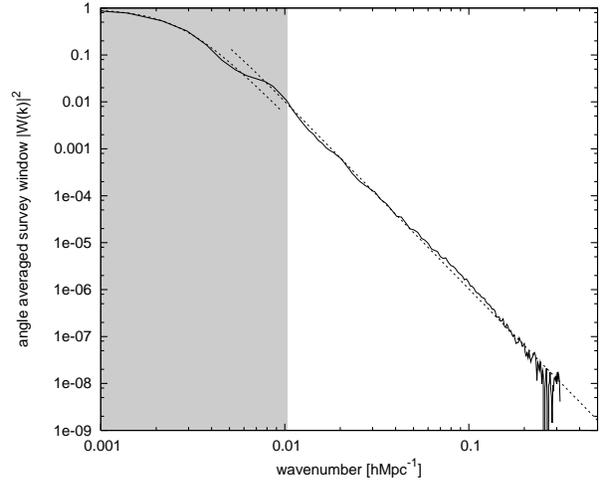}
\caption{Isotropized survey window. Here the normalization is taken such that $|W(0)|=1$. Light gray stripe marks the region where the window is above $1\%$ of its maximum value of 1. Dashed lines show approximations discussed in the text.}
\label{fig8}
\end{figure}

Since masking in real space is equivalent to convolution in Fourier space, our measured power spectrum $\tilde{P}$ is actually a convolution of the real spectrum $P$ with a survey window (see e.g. FKP):
\begin{equation}\label{eq7}
\tilde{P}(\mathbf{k})= \int \frac{{\rm d}^3k'}{(2\pi)^3}P(\mathbf{k})|W(\mathbf{k}-\mathbf{k'})|^2\,,
\end{equation}
where 
\begin{equation}
W(\mathbf{k})=\int {\rm d}^3r \, \bar{n}(\mathbf{r}) w(\mathbf{r}) \exp(i\mathbf{k}\cdot\mathbf{r})\,,
\end{equation}
and the survey window $|W(\mathbf{k})|^2$ is normalized as follows:  
\begin{equation}
\int \frac{{\rm d}^3k}{(2\pi)^3}|W(\mathbf{k})|^2=1\,. 
\end{equation}
The angle averaged survey window $|W(k)|^2$ is plotted in Fig. \ref{fig8}. Here the core part of the window is well approximated by the functional form: 
\begin{equation}
|W(k)|^2 = \frac{1}{1+\left(\frac{k}{a}\right)^2+\left(\frac{k}{b}\right)^4}, \quad (a\simeq 0.0030, b\simeq 0.0028)\,,
\end{equation}
and asymptotic wings are close to the power law with spectral index $-4$. These approximations are shown with dashed lines in Fig. \ref{fig8}. With the gray shaded stripe we have marked the scales where $|W(k)|^2$ is above $1 \%$ of its maximum value. This stripe just serves as a rough guide to the effective width of the survey window and it is also shown in many of the following figures.

\begin{figure}
\centering
\includegraphics[width={8.cm},angle=90]
{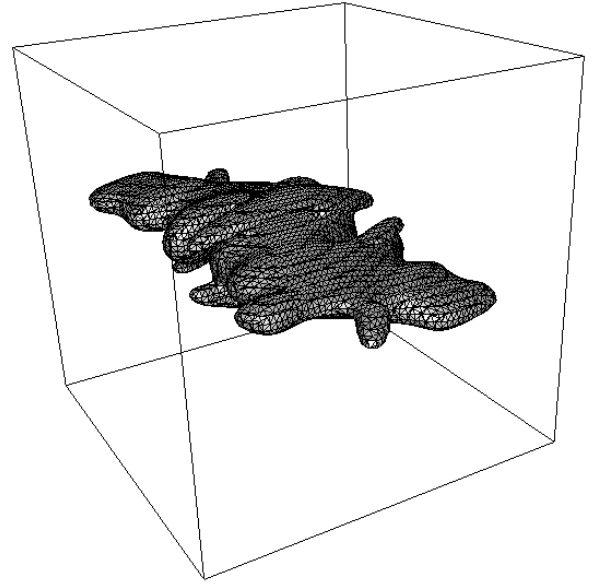}
\caption{3D survey window embedded in a box with a side length of $0.04\,h\,\mathrm{Mpc}^{-1}$. Here the isosurface corresponding to $1 \%$ of the maximum value of the window is shown. Note the symmetry of the window, $|W(\mathbf{k})|^2=|W(-\mathbf{k})|^2$, as expected when taking a modulus of the Fourier transform of a real 3D scalar function.} 
\label{fig9}
\end{figure}

Since the survey geometry of the analyzed SDSS LRG sample is far from being spherically symmetric, an isotropized window in Fig. \ref{fig8} gives only a poor representation of the true 3D window, which is displayed as an isosurface corresponding to the isovalue of $0.01$ in Fig. \ref{fig9}. 

\begin{figure}
\centering
\includegraphics[width=\plotwd]
{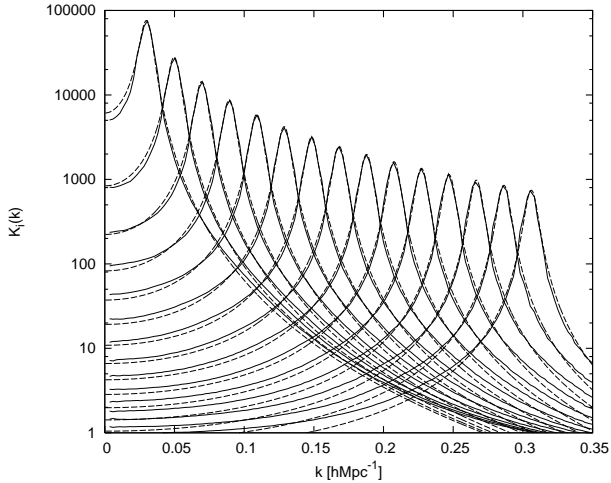}
\caption{Coupling kernels $K_i(k)\equiv K(k_i,k)$ for the power spectrum bins $k_i$ shown in Fig. \ref{fig5}. Numerically evaluated kernels are shown with solid lines. The dashed lines correspond to the fitting functions given in Appendix \ref{appd}.}
\label{fig10}
\end{figure}

In order to compare theoretical models to the measured power spectrum we have to take into account the smearing effects caused by the survey window. Using Eq. (\ref{eq7}) we can express an isotropized power spectrum as:
\begin{equation}\label{eq10}
\tilde{P}(k) = \int \frac{{\rm d}\Omega_{\mathbf{k}}}{4\pi}\tilde{P}(\mathbf{k}) = \int {\rm d}k'\,k'^2P(k')K(k',k) \,,
\end{equation}
where the coupling kernels \footnote{We prefer to use ``coupling kernels'' instead of the more common ``window functions'' since the word ``window'' has already been used to mean the modulus square of the Fourier transform of the weighted survey volume.}:
\begin{equation}
K(k',k) = K(k,k') = \frac{1}{2\pi^2}\int \frac{{\rm d}\Omega_{\mathbf{k}}}{4\pi} \int \frac{{\rm d}\Omega_{\mathbf{k'}}}{4\pi}|W(\mathbf{k}-\mathbf{k'})|^2\,.
\end{equation}
Numerically evaluated coupling kernels along with the analytical approximations (see Appendix \ref{appd}) for the analyzed galaxy sample are presented in Fig. \ref{fig10}. Here the solid lines correspond to the numerical results and the dashed ones represent an analytical approximation. We have used the notation $K_i(k)\equiv K(k_i,k)$ where $k_i$ denote the central values of the power spectrum bins shown in Fig. \ref{fig5}.

\begin{figure}
\centering
\includegraphics[width=\plotwd]
{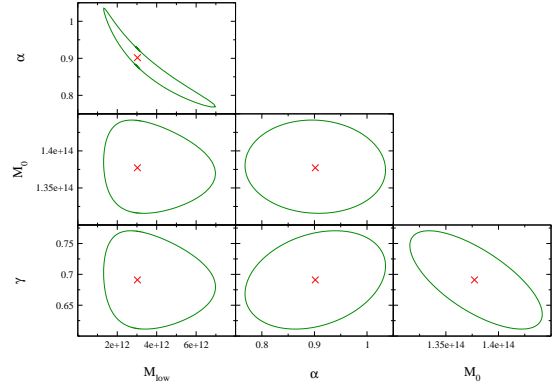}
\caption{$1\sigma$ error contours for the free model parameters. Best fit parameter values are marked with crosses.}
\label{fig11}
\end{figure}

\section{Model spectra}\label{sec6}
It is well known that redshift space distortions and nonlinear effects modify simple linear spectra. In order to treat these effects we make use of the very successful analytical model -- the Halo Model. For a nice review we refer the reader to \citet{2002PhR...372....1C} (see also \citealt{2000MNRAS.318..203S}). The details of the model we use are presented in Appendix \ref{appc}. The model introduces four free parameters: $M_{\mathrm{low}}$, $\alpha$, $M_0$ and $\gamma$. Here $M_{\mathrm{low}}$ is the lower cutoff of the halo mass i.e. below that mass halos are assumed to be ``dark''. $\alpha$ and $M_0$ are the parameters of the mean of the halo occupation distribution $\langle N|M \rangle$, which gives the average number of galaxies per halo with mass $M$. We take $\langle N|M \rangle$ to be a simple power law:
\begin{equation}
\langle N|M \rangle = \left ( \frac{M}{M_0} \right )^\alpha\,.
\end{equation}
The last parameter, $\gamma$, is the amplitude factor for the virial velocities of galaxies inside dark matter halos. One dimensional velocity dispersion of the galaxies inside a halo with mass $M$ is taken to follow the scaling of the isothermal sphere model:
\begin{equation}
\sigma = \gamma \sqrt {\frac{GM}{2R_\mathrm{vir}}}\,,
\end{equation}
where $R_\mathrm{vir}$ is the virial radius of the halo.

For the model fitting we have used Levenberg-Marquardt method as described in \citet{1992nrfa.book.....P} with modifications (described in Appendix \ref{appe}) that allow us to incorporate correlations between the data points. As the input data we take the power spectrum estimates given in Fig. \ref{fig5}. The covariance matrix used is the one shown in the middle row of Fig. \ref{fig7}. We also perform fits where we use one additional power spectrum bin on a larger scale (not shown in Fig. \ref{fig5}). All of this data is given in a tabular form in Appendix \ref{appg}. The transfer functions needed for the linear spectra are taken from \citet{1998ApJ...496..605E}. There the authors also provide transfer function fits where the baryonic acoustic oscillations have been removed. We use these ``smoothed out'' transfer functions in order to assess the significance of the oscillatory features we see in the data. Throughout this paper we have kept cosmology fixed to the best fit WMAP ``concordance'' model \citep{2003ApJS..148..175S}. The implications for the cosmology, and especially for the dark energy equation of state parameter, are planned to be worked out in the future paper.

As the cosmology is kept fixed, we have only four free parameters. In order to eliminate some of the degeneracies between the parameters we have imposed one additional constraint. Namely, we have demanded that the resulting number of galaxies should agree with the one that is observed with the relative error of $1\%$ i.e. $(51,763 \pm 518)$\footnote{The $1 \sigma$ Poisson error in this case would be $228$. The large-scale structure amplifies the variability in the number of objects and a factor of a few increase above the Poissonian case seems to be reasonable.}. The resulting $1\sigma$ error ``ellipses'' for the free parameters are shown in Fig. \ref{fig11}. The ``ellipses'' appear deformed since instead of $M_{\mathrm{low}}$ and $M_0$ we have fitted $\log(M_{\mathrm{low}})$ and $\log(M_0)$. With crosses we have marked the best fit values: $M_{\mathrm{low}} \simeq 3\cdot 10^{12} h^{-1}M_{\odot}$, $\alpha \simeq 0.9$, $M_0 \simeq 1.4 \cdot 10^{14} h^{-1}M_{\odot}$ and $\gamma \simeq 0.7$. The model spectra corresponding to these best fit parameters are shown in Figs. \ref{fig4} and \ref{fig5}. In both figures we have also given the simple linear spectra multiplied by the square of the bias parameter $b=1.95$. In Fig. \ref{fig4} we have additionally demonstrated the effect of the window convolution. There the dashed lines correspond to the unconvolved case. In Fig. \ref{fig5} along with the ``wiggly'' spectra we have shown their ``smoothed-out'' counterparts. Using all the 16 power spectrum bins (the 1st not shown in Fig. \ref{fig5}) plus an additional constraint on the total number of galaxies, resulting in $17 - 4 = 13$ independent degrees of freedom, we obtain $\chi^2$ values of $8.8$ and $19.9$ for the ``wiggly'' and ``smoothed''\footnote{The best fit $M_{\mathrm{low}}$, $\alpha$, $M_0$ and $\gamma$ for the ``smoothed'' models differ slightly from the values quoted above for the ``wiggly'' spectra.} models, respectively. So the models with oscillations are favored by $3.3 \sigma$ over their ``smoothed-out'' counterparts.\footnote{Dropping the first power spectrum bin the obtained $\chi^2$ values are $5.0$ and $16.5$. ~$5.0$ is an anomalously low value of $\chi^2$ for $12$ degrees of freedom. (One would expect $\chi^2 \simeq 12 \pm 5$.) In fact, if we would have used the simple FKP covariance matrix instead of the one obtained from the mock catalogs, the resulting $\chi^2$ values would be even lower: $2.9$ and $8.5$, respectively. This might hint that the 2nd order Lagrangian approach, although very successful, might still have problems of capturing some extra mode-mode couplings.} Since both models have the same number of free parameters, and if additionally the assumption of Gaussianity is valid, the Bayesian approach should also give similar results. Actually, Bayesian results should favor ``wiggly'' models even more, since prior weight for these should probably be taken higher (assuming the knowledge of the other experimental results).

\section{Determination of the acoustic scale}

\begin{figure}
\centering
\includegraphics[width=\plotwd]
{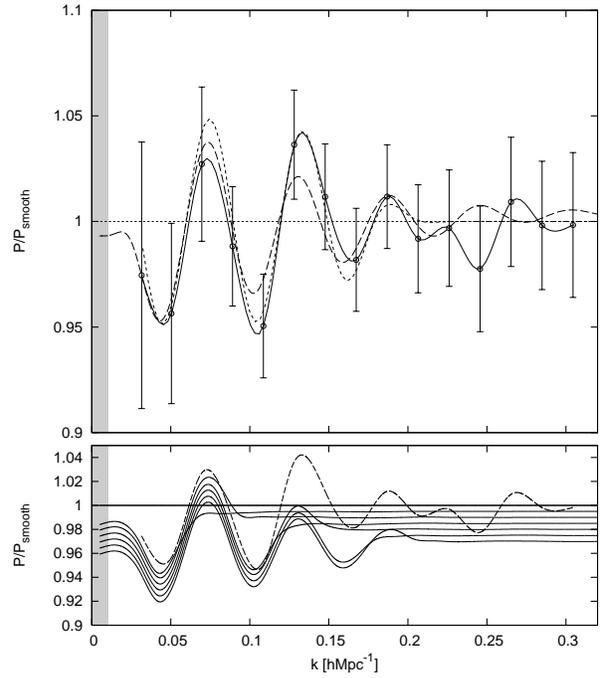}
\caption{Upper panel: Power spectrum from Fig. \ref{fig5} divided by the best fitting ``smoothed'' spectrum. Solid line shows a cubic spline fit to the data points and long-dashed line corresponds to the best ``wiggly'' model. The short-dashed line represents the most favorable fit from the parametric family of Eq. (\ref{eq8}). Lower panel: Various input power spectra used to calculate the two-point correlation function. The dashed line is the cubic spline fit from the upper panel. The solid lines represent a transition sequence from the best fitting ``wiggly'' model to the best ``smoothed'' model. In each step we have erased more and more oscillatory features. For clarity slight vertical shifts have been introduced.}
\label{fig12}
\end{figure}

To measure the scale of the acoustic oscillations we divide the spectrum shown in Fig. \ref{fig5} with the best fitting ``smoothed'' spectrum. The result of this procedure is given in the upper panel of Fig. \ref{fig12}. There the solid line shows a cubic spline fit to the data points and the long-dashed line corresponds to the best fitting model spectrum also shown in Fig. \ref{fig5}. The above data is fitted with a parametric form:
\begin{equation}\label{eq8}
f(x)=1+c_1\cdot\sin(c_2\cdot x)\exp\left[\left(-\frac{x}{c_3}\right)^{c_4}\right]\,.
\end{equation}
Again we use the Levenberg-Marquardt method with the data covariance matrix obtained from mock catalogs. After marginalizing over the other parameters we find the best fitting value of $(105.4 \pm 2.3)\,h^{-1}\,\mathrm{Mpc}$ for the parameter $c_2$. \footnote{Here and in the following all the errors refer to the 1-$\sigma$ level. Values for the other parameters are as follows: $c_1 = (4.9 \pm 2.1)\cdot 10^{-2}$, $c_3 = (0.176 \pm 0.023)\,h\,\mathrm{Mpc}^{-1}$, $c_4 = (7 \pm {}^{17}_{\phantom{1}7})$.} The best fitting member of the parametric family in Eq. (\ref{eq8}) is shown with short-dashed lines in the upper panel of Fig. \ref{fig12}. Using FKP covariance matrix instead gives an acoustic scale of $(105.4 \pm 2.8)\,h^{-1}\,\mathrm{Mpc}$.

The sinusoidal modulation in the power spectrum is a pure consequence of the adiabatic initial conditions. By relaxing this assumption %(e.g. allowing for some isocurvature component) 
and fitting with a more general functional form: 
\begin{equation}
f(x)=1+c_1\cdot\sin(c_2\cdot x + c_3)\exp\left[\left(-\frac{x}{c_4}\right)^{c_5}\right]\,.
\end{equation}
instead, we get the following value for the acoustic scale: $(103.0 \pm 7.6)\,h^{-1}\,\mathrm{Mpc}$. In case of the FKP covariance matrix the corresponding value is $(103.1 \pm 9.1)\,h^{-1}\,\mathrm{Mpc}$. 

\citet{2005ApJ...633..560E} determine various distance scales (like $D_v$, which is a certain mixture of the comoving distances along and perpendicular to the line of sight (see their Eq.(2))) and their ratios, using SDSS LRGs in combination with the constraints from other cosmological sources. The typical relative accuracy of these measurements is $\sim 4 \%$, which might seem to be significantly poorer than the accuracy of the acoustic scale measurement, $(105.4 \pm 2.3)\,h^{-1}\,\mathrm{Mpc}$ i.e. $\sim 2\%$, presented in this paper. This apparent inconsistency can be attributed to the fact that in our analysis, as stated above, we have kept the cosmology fixed to the WMAP ``concordance'' model, whereas \citet{2005ApJ...633..560E} estimates include the extra uncertainties due to the imperfect knowledge of the various cosmological parameters. Of course, the given length of the acoustic scale, $(105.4 \pm 2.3)\,h^{-1}\,\mathrm{Mpc}$, can be easily transformed in order to accommodate other preferences for the background cosmology. We also note that the use of the parametric form in Eq. (\ref{eq8}) might be too restrictive, since the acoustic modulation in the case of adiabatic models can be only approximately described as a damped sinusoidal wave \citep{1998ApJ...496..605E}. For this reason the given sound horizon constraint should not be used in cosmological parameter studies. Instead one should directly use the measured power spectrum in combination with the parametrized models that are physically well motivated. 

\section{Correlation function analysis}

\begin{figure}
\centering
\includegraphics[width=\plotwd]
{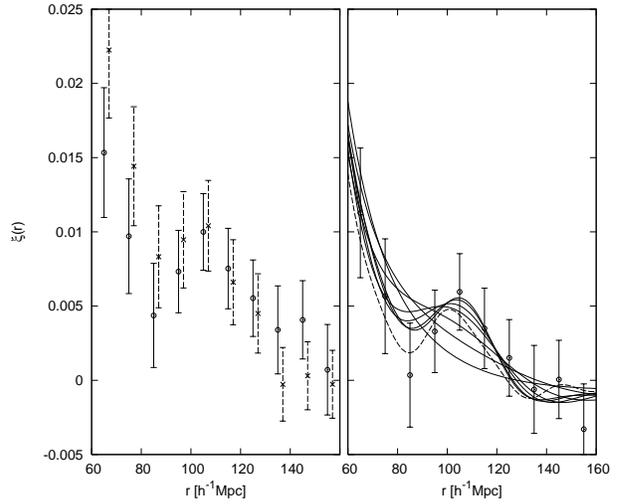}
\caption{Left panel: Two-point correlation functions as determined in this paper (circles with solid lines) and by \citet{2005ApJ...633..560E}. Right panel: Correlation functions corresponding to the models shown in the lower panel of Fig. \ref{fig12} in comparison to the one obtained directly from the data. Here all the data points have been lowered by $0.0035$.}
\label{fig13}
\end{figure}

We determine the two-point correlation function of the SDSS LRGs using the edge-corrected estimator given by \citet{1993ApJ...412...64L}:
\begin{equation}\label{eq9}
\xi (r) = \frac{DD - 2DR + RR}{RR}\,,
\end{equation}
which has minimal variance for a Poisson process. Here DD, DR and RR represent the respective normalized data-data, data-random and random-random pair counts in a given distance range. Random catalogs were generated with $25$ times the number of objects in the main catalog. We calculated correlation function for $10$ bins ($r_i,\,i=1\ldots10$) in the pair distance range of $60 \ldots 160\,h^{-1}\,\mathrm{Mpc}$. The errors were estimated by a ``jackknife'' technique. For this purpose we divided the full sample into $22$ separate regions each containing $\sim 2350$ galaxies (see Fig. \ref{fig3}). The two-point function was calculated $22$ times, each time omitting one of the regions. Denoting the resulting estimates as $\xi_j(r_i),\,(j=1\ldots22)$, the ``jackknife'' estimate for the variance reads as (see e.g. \citealt{1993stp..book.....L}):
\begin{eqnarray}
\sigma^2_{\xi}(r_i) &=& \frac{N-1}{N}\sum \limits_{j=1}^N \left[\xi_j(r_i) -\bar{\xi}(r_i)\right]^2\,,\\
\bar{\xi}(r_i) &=& \frac{1}{N}\sum \limits_{j=1}^N \xi_j(r_i)\,, 
\end{eqnarray}   
where in our case $N=22$. The results of this calculation are presented in the left panel of Fig. \ref{fig13}. With the crosses and dashed-line errorbars we have also shown the two-point function as determined by \citet{2005ApJ...633..560E}. We see that in general our results agree reasonably well with their calculations.

It would be interesting to study how the oscillations in the observed power spectrum transform into the peak in the two-point correlation function seen at the scale of $\sim  110 \,h^{-1}\,\mathrm{Mpc}$. For this purpose we use the cubic spline fit shown in Fig. \ref{fig12} and extend it outside of the observed range by smoothly joining it to the power spectrum of the best fitting ``smoothed-out'' model. The correlation function is now simply calculated as the Fourier transform of the power spectrum. \footnote{To be precise, in redshift space the two-point correlation function and power spectrum are not anymore exact Fourier transforms of each other. Nevertheless, we think that this simplified exercise is still useful. Also, as the correlation function estimator in Eq. (\ref{eq9}) is an edge-corrected estimator, we use an unconvolved model spectra here.} The resulting correlation function is plotted with a dashed line in the right panel of Fig. \ref{fig13}. To study the significance of the oscillatory features in the power spectrum in relation to the observed peak in the correlation function, we have calculated correlation functions for several models that have oscillations ``switched off'' at various scales. The spectra of these models are shown with solid lines in the lower panel of Fig. \ref{fig12}, where for the sake of clarity we have introduced slight vertical shifts between the curves, so that the scales where the transition to the featureless spectrum takes place, are easily visible. The corresponding correlation functions are given with solid lines in the right hand panel of Fig. \ref{fig13}. As expected, we see how the peak in the correlation function is getting broader and also decreasing in amplitude as we erase more and more features in the power spectrum. This clearly demonstrates the importance of many of the up-downs in the power spectrum to produce a relatively sharp feature in the two-point correlation function. 

In order to achieve good agreement we have lowered all the data points by $0.0035$ in the right hand panel of Fig. \ref{fig13}. Similar shifts were also suggested in \citet{2005ApJ...633..560E} in order to get better match to the theoretical models. A $0.0035$ shift in $\xi$ translates to the $0.175 \%$ shift in the mean density. Thus, if one wishes to determine the amplitude of the correlation function correctly at those large scales, one has to determine the survey selection function with a very high precision, which in practice is very difficult to achieve. By using model spectra that have more large scale power than the ``concordance'' cosmology predicts (as might be suggestive from Fig. \ref{fig4}), we are in fact able to match the amplitude of the correlation function without any additional vertical shifts. Here we try to avoid making any definite conclusions. The behavior of the power spectrum on the largest scales is an extremely interesting topic on its own and there exist much better methods than the direct Fourier approach to investigate these issues (see e.g. \citealt{1998ApJ...499..555T}).       

\section{Comparison with the other surveys}

\begin{figure}
\centering
\includegraphics[width=\plotwd]
{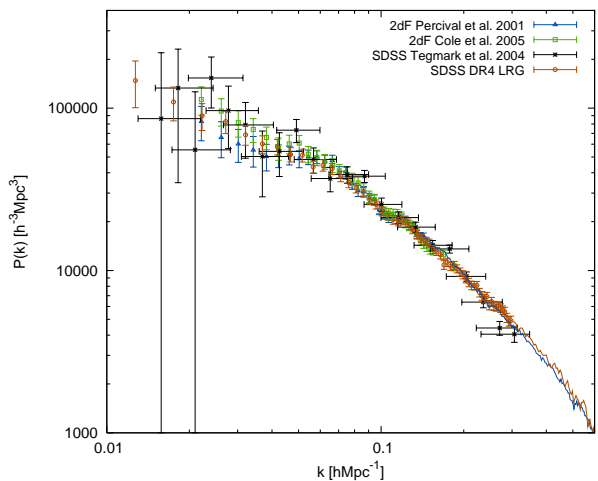}
\caption{The comparison of spectra from different surveys.}
\label{fig14}
\end{figure}

\begin{figure}
\centering
\includegraphics[width=\plotwd]
{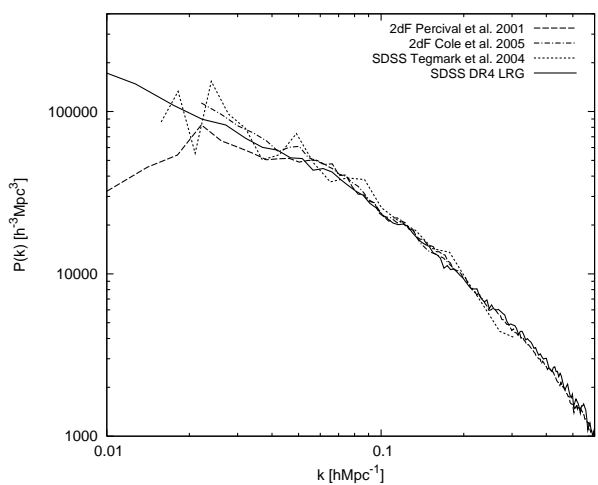}
\caption{The same as Fig. \ref{fig14} with the errorbars omitted.}
\label{fig15}
\end{figure}

In this section we compare our power spectrum measurements with the ones obtained by \citet{2001MNRAS.327.1297P} and \citet{2005MNRAS.362..505C} for the 2dF redshift survey and by \citet{2004ApJ...606..702T} for the SDSS main galaxy sample. The results of this comparison are provided in Figs. \ref{fig14} and \ref{fig15}. Since Fig. \ref{fig14} is extremely busy, we have also given a variant of it where we have omitted the errorbars. The amplitudes of the SDSS main and 2dF spectra have been freely adjusted to match the clustering strength of the SDSS LRGs. The corresponding bias parameters with respect to the SDSS LRGs are $0.53$, $0.61$ and $0.50$ for the 2dF sample analyzed by \citet{2001MNRAS.327.1297P}, for the one analyzed by \citet{2005MNRAS.362..505C}, and for the SDSS main sample, respectively. \citet{2001MNRAS.327.1297P} also provide power spectrum measurements for $k \gtrsim 0.15\,h\,\mathrm{Mpc}^{-1}$ but without errorbars. These small-scale measurements along with our SDSS LRG results are shown with solid lines in Fig. \ref{fig14}.

In general the shapes of the spectra agree remarkably well. Of course one has to keep in mind that here, with the only exception of \citet{2004ApJ...606..702T} results, the power spectrum bins are highly correlated. Also \citet{2004ApJ...606..702T} measurements are corrected for the redshift space distortions.   
\section{Discussion and Conclusions}
In this paper we have calculated the redshift-space power spectrum of the SDSS DR4 LRG sample, finding evidence for a series of acoustic features down to the scales of $\sim 0.2\,h\,\mathrm{Mpc}^{-1}$. It turns out that models with the baryonic oscillations are favored by $3.3 \sigma$ over their ``smoothed-out'' counterparts without any oscillatory behavior. Using the obtained power spectrum we predict the shape of the spatial two-point correlation function, which agrees very well with the one obtained directly from the data. Also, the directly calculated correlation function is consistent with the results obtained by \citet{2005ApJ...633..560E}. We have made no attempts to put constraints on the cosmological parameters, rather we have assumed in our analysis the ``concordance'' cosmological model. The derived acoustic scale $(105.4 \pm 2.3)\,h^{-1}\,\mathrm{Mpc}$ agrees well with the best-fit WMAP ``concordance'' model prediction of $\simeq 106.5 \,h^{-1}\,\mathrm{Mpc}$. 

The existence of the baryonic features in the galaxy power spectrum is very important, allowing one (in principle) to obtain Hubble parameter $H$ and angular diameter distance $d_A$ as a function of redshift, this way opening up a possibility to constrain properties of the dark energy \citep{2003PhRvD..68f3004H}. The currently existing biggest redshift surveys, which are still quite shallow, do not yet provide enough information to carry out this project fully. On the other hand, it is extremely encouraging that even with the current generation of redshift surveys we are already able to see the traces of acoustic oscillations in the galaxy power spectrum, showing the great promise for the dedicated future surveys like K.A.O.S. We have seen that acoustic features seem to survive at mildly nonlinear scales ($k \gtrsim 0.1\,h\,\mathrm{Mpc}^{-1}$), which is in agreement with the results of the recent N-body simulations \citep{2005Natur.435..629S,2005ApJ...633..575S}. In order to fully exploit available information one needs a complete understanding of how nonlinear effects influence these features. Nonlinear bias and redshift space distortions also add extra complications. In general redshift-space distortions, biasing and nonlinear evolution do not create any oscillatory modulation in the power spectrum and so acoustic features should be readily observable. 
%On the other hand, only the effects that change the amplitude of the spectrum (e.g. redshift-space distortions on large scales) are easily factored out, but the ones involving stretching/compressing of the spatial scales do start to interfere with the cosmological distortions and so inhibit our ability to draw conclusions about the underlying cosmological model.%
 So far there have been only a few works studying these important issues (e.g.  \citet{2005Natur.435..629S,2005ApJ...633..575S,2005APh....24..334W}) and probably it is fair to say that currently we really do not have a full theoretical description of them. In our paper we have modeled the above mentioned effects using the results from the 2nd order Lagrangian perturbation theory in combination with the Halo Model. Although these models are very successful in capturing many important aspects of the structure formation, one has to keep in mind that they are still approximations. 

The bare existence of the baryonic oscillations in the galaxy power spectrum tells us something important about the underlying cosmological model and the mechanism of the structure formation. First, it confirms the generic picture of the gravitational instability theory where the structure in the Universe is believed to be formed by the gravitational amplification of the small perturbations layed down in the early Universe. Under the linear gravitational evolution all the density fluctuation modes evolve independently i.e. all the features in the power spectrum will be preserved. And certainly, we are able to identify features in the low redshift galaxy power spectrum that correspond to the fluctuations seen in the CMB angular power spectrum (which probes redshifts $z \sim 1100$), providing strong support for the above described standard picture of the structure formation. Actually, we can also probe scales that are unaccessible for the CMB studies due to the strong damping effects and steeply rising influence of the secondary anisotropies, reaching effectively the wavenumbers that correspond to the 6th-7th peak in the CMB angular power spectrum. Second, the ability to observe baryonic features in the low redshift galaxy power spectrum demands rather high baryonic to total matter density ratio. In \citet{2003A&A...412...35B} it has been shown that it is possible to fit a large body of observational data with an Einstein--de Sitter type model if one adopts low value for the Hubble parameter and relaxes the usual assumptions about the single power law initial spectrum. In the light of the results obtained in our paper these models are certainly disfavored due to the fact that the high dark matter density completely damps the baryonic features. And finally, purely baryonic models are also ruled out since for them the expected acoustic scale would be roughly two times larger than observed here \footnote{For a clear discussion of this see Daniel Eisenstein's home page http://cmb.as.arizona.edu/ $\sim$eisenste/acousticpeak/}. So the data seems to demand a weakly interacting nonrelativistic matter component and all the models that try to replace this dark matter component with something else e.g. modifying the laws of gravity might have severe difficulties to fit these new observational constraints.  

\acknowledgements{I thank Rashid Sunyaev for valuable comments on the manuscript. Also I am grateful to the referee for helpful comments and suggestions. I acknowledge the International Max-Planck Research School on Astrophysics for a graduate fellowship and the support provided through Estonian Ministry of Education and Recearch project TO 0062465S03 and and ESF grant 5347.\\
    Funding for the creation and distribution of the SDSS Archive has been provided by the Alfred P. Sloan Foundation, the Participating Institutions, the National Aeronautics and Space Administration, the National Science Foundation, the U.S. Department of Energy, the Japanese Monbukagakusho, and the Max Planck Society. The SDSS Web site is http://www.sdss.org/.

    The SDSS is managed by the Astrophysical Research Consortium (ARC) for the Participating Institutions. The Participating Institutions are The University of Chicago, Fermilab, the Institute for Advanced Study, the Japan Participation Group, The Johns Hopkins University, the Korean Scientist Group, Los Alamos National Laboratory, the Max-Planck-Institute for Astronomy (MPIA), the Max-Planck-Institute for Astrophysics (MPA), New Mexico State University, University of Pittsburgh, University of Portsmouth, Princeton University, the United States Naval Observatory, and the University of Washington.
}

\bibliographystyle{aa}
\bibliography{references}

\appendix

\section{Test problem}\label{appa}
\begin{figure}
\centering
\includegraphics[width=\plotwd]
{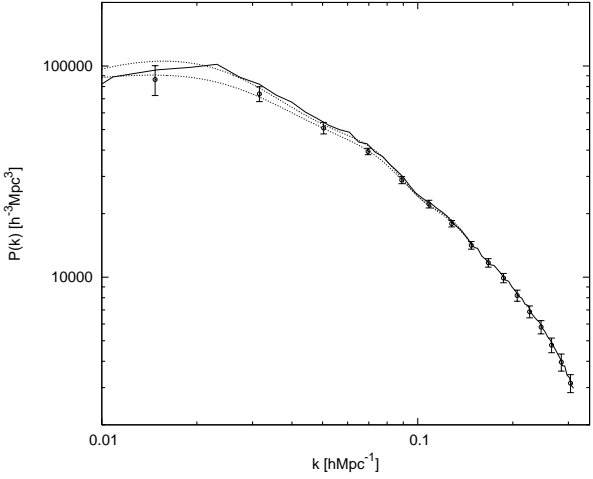}
\caption{Power spectra of galaxy clusters from the $z=0$ Hubble Volume simulation box. The solid line represents the spectrum for the full sample of $1,560,995$ clusters. The circles with errorbars denote the recovered spectrum from the $50$ mock catalogs having similar selection effects to the analyzed SDSS LRG sample. The dotted lines demonstrate the convolution effect of the survey window on the best fitting model spectrum.}
\label{fig_a1}
\end{figure}

Here we present one test for our power spectrum calculation software \footnote{Further tests can be found in \citet{astro-ph/0505441}}. As the input we use the $z=0$ cluster catalog of the VIRGO Hubble Volume simulations\footnote{http://www.mpa-garching.mpg.de/Virgo/}, which covers the comoving volume of $3000\,h^{-3}\,\mathrm{Mpc}^3$ and contains $1,560,995$ clusters above the mass limit of $6.75\cdot 10^{13} h^{-1}M_{\odot}$. The average bias parameter of this catalog is $b = 1.9$, which is comparable to the SDSS LRG value of $b = 1.95$. The power spectrum of the full sample is shown in Fig. \ref{fig_a1} with a solid line. Here for clarity we have not shown the errorbars, which are rather small for a sample of that size. Out of the full sample we generate $50$ mock catalogs that have the same radial and angular selection functions as the SDSS LRG sample analyzed in this paper (see Figs. \ref{fig2} and \ref{fig3}). The mean number of objects in the resulting catalogs is $\sim 18,500$ i.e. the number density is roughly one third of the spatial density of the SDSS LRGs. Observer's location and pointing angles are taken randomly for each of the catalogs. The mean recovered power spectrum with $1 \sigma$ errorbars is shown in Fig. \ref{fig_a1}. We see that the power spectrum of the underlying sample is recovered very well. On the largest scales there are some deviations, which can be explained as being caused by the smearing effect of the survey window. This is demonstrated by the dotted lines, where the lower/upper curve corresponds to the model spectrum with/without survey window convolution applied.

\section{Mock catalogs}\label{appb}
We build mock catalogs for the SDSS LRG by a 3 step procedure:
\begin{enumerate}
\item Generation of the density field using an optimized 2nd order Lagrangian perturbation calculation (2LPT).
\item Poisson sampling of the generated density field with the intensity of the process adjusted so, as to end up with a galaxy sample that has a clustering strength enhanced by a factor $b^2$ with respect to the underlying field, and a number density equal to the observed LRG sample density at the minimal used redshift of $0.16$ (see Fig. \ref{fig2}).  
\item Extraction of the final catalog by applying the radial and angular selection function as given in Figs. \ref{fig2} and \ref{fig3}, respectively.
\end{enumerate}

In contrast to the Eulerian perturbation theory, where one does a perturbative expansion of the density contrast field, Lagrangian approach considers an expansion of the particle trajectories (see e.g. \citealt{1993MNRAS.264..375B,1995A&A...296..575B,1995PhR...262....1S,2002PhR...367....1B}). Here the central quantity is the displacement field $\mathbf{\Psi} (\mathbf{q})$, which relates particle's initial comoving position (Lagrangian position) $\mathbf{q}$ to its final Eulerian position $\mathbf{x}$:
\begin{equation}\label{eq_a1}
\mathbf{x} = \mathbf{q} + \mathbf{\Psi} (\mathbf{q})\,.
\end{equation}
It turns out that due to the decay of the rotational perturbation modes in the expanding Universe each order of the perturbation theory displacement field separates into a time-dependent and a Lagrangian position dependent factors \citep{1997GReGr..29..733E}. The position dependent part, due to its irrotational nature can be given as a gradient of a scalar potential. As a result, one can expand the displacement field as follows:
\begin{equation} \label{eq_a2}
\mathbf{\Psi} (\mathbf{q}) = D_1 \nabla_q \phi^{(1)} + D_2 \nabla_q \phi^{(2)}\,.
\end{equation}
Here the 1st term describes the classical Zeldovich approximation \citep{1970A&A.....5...84Z}. The time independent potentials $\phi^{(1)}$ and $\phi^{(2)}$ are found from the Poisson equations:
\begin{equation}
\Delta \phi^{(1)}(\mathbf{q}) = -\delta(\mathbf{q})
\end{equation}
and
\begin{equation}
\Delta \phi^{(1)}(\mathbf{q}) = \frac{1}{2}\sum \limits_{i} \sum
\limits_{j} \left ( \phi^{(1)}_{,ii}(\mathbf{q})\phi^{(1)}_{,jj}(\mathbf{q})
- \phi^{(1)}_{,ij}(\mathbf{q})\phi^{(1)}_{,ji}(\mathbf{q})\right )\,,
\end{equation}
where $_{,i}$ denotes the partial derivative with respect to the Lagrangian coordinate $q_i$. $\delta(\mathbf{q})$ is the initial density contrast. We generate $\delta(\mathbf{q})$ using the standard Zeldovich approximation on a regular cubical grid. 

$D_1$ in Eq. (\ref{eq_a2}) is the linear growth factor. The second-order growth factor $D_2$ for flat models with a cosmological constant is to a good precision approximated as \citep{1995A&A...296..575B}:
\begin{equation}
D_2 \simeq -\frac{3}{7}\Omega_m^{-1/143}D_1^2\,.
\end{equation}
According to Eq. (\ref{eq_a1}) and (\ref{eq_a2}) the peculiar velocity field is given as:
\begin{equation}\label{eq_a3}
\mathbf{v} = D_1 f_1 H \nabla_q \phi^{(1)} + D_2 f_2 H \nabla_q \phi^{(2)}\,.
\end{equation}
Here $H \equiv \frac{\dot{a}}{a}$ and $f_i \equiv \frac{\mathrm{d} \ln D_i}{\mathrm{d} \ln a}$. For flat models with a cosmological constant logarithmic derivatives of the growth factors can be approximated as \citep{1995A&A...296..575B}:
\begin{equation}
f_1 \simeq \Omega_m^{5/9},\quad f_2 \simeq 2\Omega_m^{6/11}\,.
\end{equation} 

\begin{figure*}
\centering
\includegraphics[width={14 cm}]
{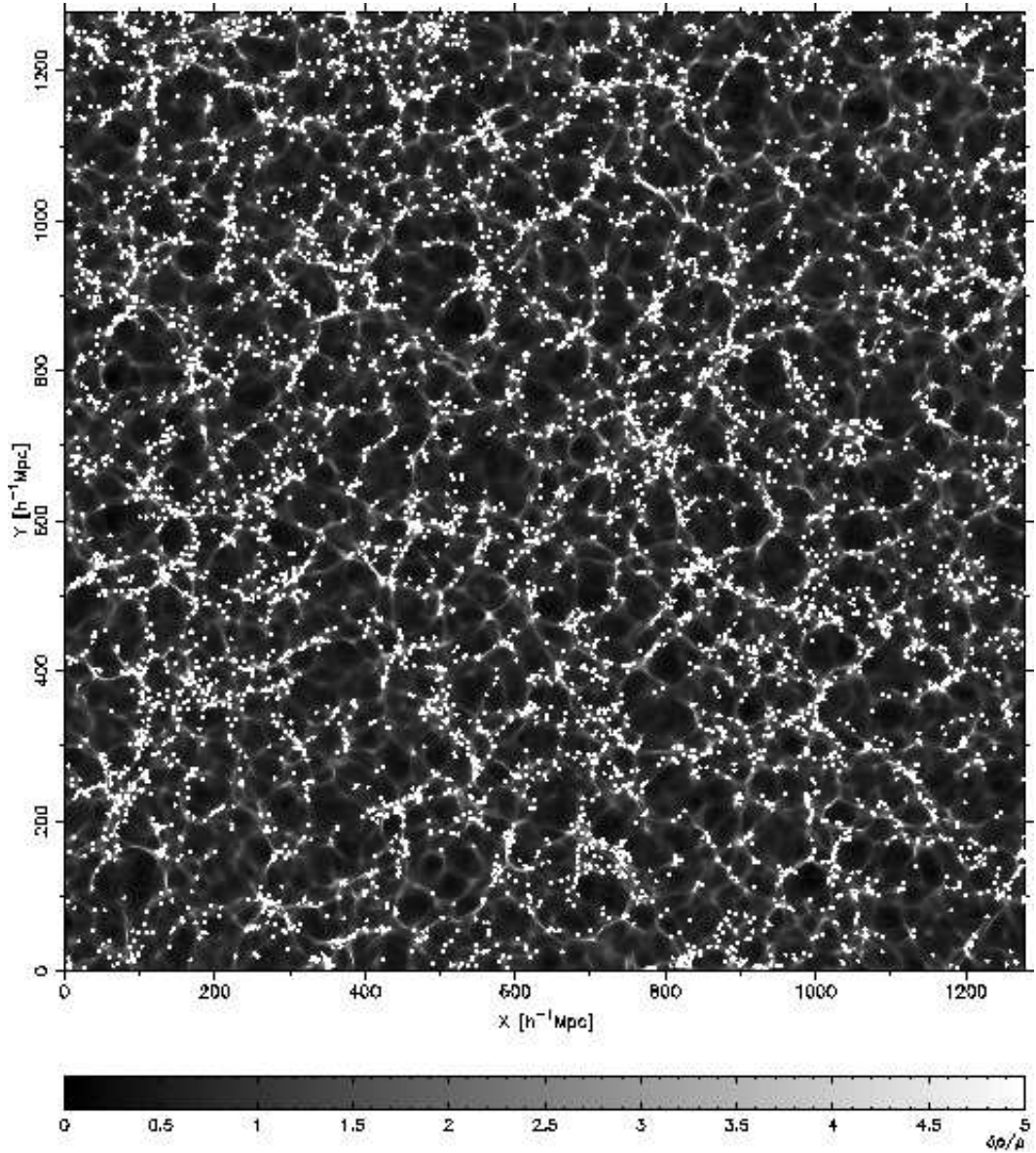}
\caption{A $25\,h^{-1}\,\mathrm{Mpc}$ thick slice through a $1280\,h^{-1}\,\mathrm{Mpc}$ computational box. A gray scale image represents the underlying density field obtained by the optimized 2LPT approach. White dots mark the positions of the ``galaxies'' generated by the Poisson sampler.}
\label{fig_a2}
\end{figure*}

Lagrangian perturbative approach works fine up to the 1st shell-crossing. After that the formed caustic structures will start to be wiped out, since the particles just keep on moving without noticing the gravitational pull of the dense sheets/filaments. It is possible to cure this problem significantly by filtering out the small-scale Fourier modes. This is what is meant by the ``optimization''. The method applied to the 1st order Lagrangian perturbation calculation is known as the truncated Zeldovich approximation (e.g. \citealt{1993MNRAS.260..765C,1994MNRAS.269..626M,1996MNRAS.278..953W}). \citet{1996MNRAS.278..953W} suggest to remove the small-scale power by applying a Gaussian $k$-space filter with a characteristic smoothing scale $k_{gs}$ to the initial density field. Thus the power spectrum of the filtered field is given by:
\begin{equation}
P_{\mathrm{optimized}}(k) = P(k)\exp \left (-\frac{k^2}{k^2_{gs}} \right )\,.
\end{equation}
They recommend the value $k_{gs}\simeq 1.2 k_{nl}$, where the nonlinearity scale $k_{nl}$ is defined as: 
\begin{equation}
\frac{D_1^2}{(2\pi)^3} \int \limits_0^{k_{nl}}\mathrm{d}^3k\,P(k) = 1\,.
\end{equation}
Although they studied only models with $\Omega_m=1$, it has been later shown by \citet{1998ApJ...507L...1H} that this ``recipe'' performs well for arbitrary Friedmann-Lema\^itre-Robertson-Walker models. 

In our calculations we assume the WMAP ``concordance'' cosmology \citep{2003ApJS..148..175S}. Linear power spectrum is taken from \citet{1998ApJ...496..605E}. We build 2LPT density field on a $256^3$-grid with $5\,h^{-1}\,\mathrm{Mpc}$ cell size using the same number of particles as the number of grid cells. \footnote{Due to the rather big cell size the truncation of the initial spectrum has a rather mild effect.} Four copies of this box are combined to form a bigger $2560 \times 2560 \times 1280 \, h^{-3}\,\mathrm{Mpc}^3$ volume. Out of that big box a sample of ``galaxies'' is selected with a radial number density as given in Fig. \ref{fig2} and with an angular mask presented in Fig. \ref{fig3}. The parameters of the Poisson sampler \footnote{We use a simple model where the intensity of the inhomogeneous Poisson process is linearly related to the underlying density field.} are tuned to give a sample with a bias parameter $b \simeq 2$ in agreement with the observed value for the SDSS LRG sample. The redshift-space catalog is built by altering the radial distances of the ``galaxies'' by $\mathbf{v}_r/H_0$, where $\mathbf{v}_r$ is the radial component of the peculiar velocity field (see Eq. (\ref{eq_a3})) and $H_0 = 100\,h\,\mathrm{km/s/Mpc}$.

\begin{figure}
\centering
\includegraphics[width=\plotwd]
{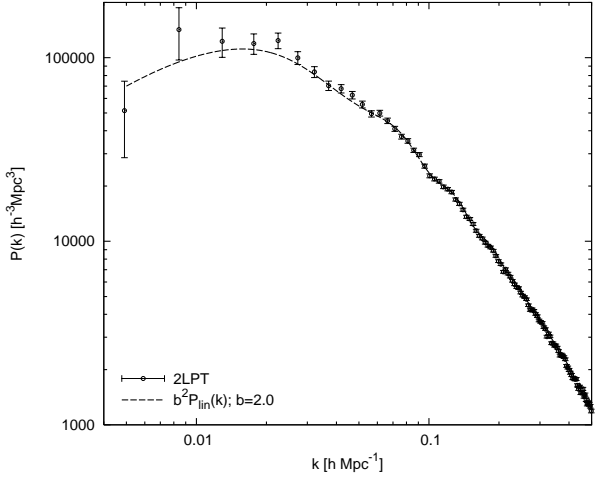}
\caption{The power spectrum of $\sim 350,000$ ``galaxies'' from the simulation box shown in Fig. \ref{fig_a2}. The solid line shows the linearly evolved input spectrum multiplied by the square of the bias parameter $b=2.0$.}
\label{fig_a3}
\end{figure}

In Fig. \ref{fig_a2} we show a $25\,h^{-1}\,\mathrm{Mpc}$ thick slice through a box with $1280\,h^{-1}\,\mathrm{Mpc}$ side length. The underlying density field is presented as a gray scale image  with white dots marking the positions of the ``galaxies''. The power spectrum of the sample of $\sim 350,000$ ``galaxies'' is shown in Fig. \ref{fig_a3} \footnote{Here as in the previous figure the SDSS LRG selection functions are not applied yet.}. We see that the shape of the spectrum is in good agreement with the linearly evolved power spectrum up to the scales of $k \sim 0.5 \,h\,\mathrm{Mpc}^{-1}$.

This approach gives us a ``galaxy'' sample that has rather realistic large-scale clustering properties. In terms of the Halo Model (see Appendix \ref{appc}) one can say that halo-halo clustering term is properly accounted for. 2LPT also gives reasonably accurate higher order correlations on quasi-nonlinear scales (e.g. \citealt{1995A&A...296..575B,2002MNRAS.329..629S}).        

\section{Power spectrum from the halo model}\label{appc}
The halo model description of the spatial clustering of galaxies is a development of the original idea by \citet{1952ApJ...116..144N}, where one describes the correlations of the total point set as arising from the two separate terms: (i) 1-halo term, that describes the correlations of galaxies populating the same halo, (ii) 2-halo term, which takes into account correlations of the galaxies occupying different halos. For a thorough review see \citet{2002PhR...372....1C}. Here we briefly give the results that are relevant to the current paper (see \citealt{2001MNRAS.325.1359S,2004MNRAS.348..250C}).

The power spectrum of galaxies in redshift space can be given as:
\begin{equation}
P(k) = P^{1h}(k) + P^{2h}(k)\,,
\end{equation}
where the 1-halo term:
\begin{eqnarray}
P^{1h}(k) & = & \int \mathrm{d}M \,n(M)\frac{\langle N(N-1)|M \rangle}{\bar{n}^2}\mathcal{R}_p(k\sigma)|u_g(k|M)|^p\,, \\
p & = &\left\{ 
\begin{array}{ll}
1 & \mathrm{\quad if \quad} \langle N(N-1)|M \rangle < 1\\
2 & \mathrm{\quad if \quad} \langle N(N-1)|M \rangle > 1
\end{array}
\right.
\end{eqnarray}
and the 2-halo term:
\begin{equation}
P^{2h}(k) = \left ( \mathcal{F}_g^2 + \frac{2}{3}\mathcal{F}_v\mathcal{F}_g + \frac{1}{5}\mathcal{F}_v^2 \right )P_\mathrm{lin}(k)\,. 
\end{equation}
Here:
\begin{eqnarray}
\mathcal{R}_p \left ( \alpha  = k \sigma \sqrt {\frac{p}{2}} \right ) & = & \frac{\sqrt{\pi}}{2}\frac{\mathrm{erf}(\alpha)}{\alpha}\,, \label{eq_b5} \\ 
\mathcal{F}_g & = & \int \mathrm{d}M \,n(M) b(M) \frac{\langle N|M \rangle}{\bar{n}}
\mathcal{R}_1(k\sigma) u_g(k|M)\,, \\
\mathcal{F}_v & = & f \cdot \int \mathrm{d}M\,n(M) b(M) \mathcal{R}_1(k\sigma) u(k|M)\,. \label{eq_b1}
\end{eqnarray}
In the above expressions $n(M)$ is the mass function and $b(M)$ halo bias parameter. We calculate them using a prescription by \citet{1999MNRAS.308..119S} and \citet{2001MNRAS.323....1S}. $\bar{n}$ represents the mean number density of galaxies:
\begin{equation}
\bar{n} = \int \mathrm{d}M\, n(M) \langle N|M \rangle\,. 
\end{equation}
We take the mean of the halo occupation distribution in the following form:
\begin{equation}\label{eq_b2}
\langle N|M \rangle = \left ( \frac{M}{M_0} \right )^\alpha\,,
\end{equation}
where $M_0$ and $\alpha$ are free parameters. The second moment is chosen as (see \citealt{2004MNRAS.348..250C}):
\begin{eqnarray}
\langle N(N-1)|M \rangle & = & \beta^2(M) \langle N|M \rangle ^2\,, \\
\beta(M) & = & \left \{ 
\begin{array}{ll}
\frac{1}{2} \log \left ( \frac{M}{10^{11} \ h^{-1}M_{\odot}} \right ) & \mathrm{\quad if \quad} M<10^{13} \  h^{-1}M_{\odot} \\
1 & \mathrm{\quad otherwise.}\label{eq_b4}
\end{array}
\right.
\end{eqnarray}
$f$ in Eq. (\ref{eq_b1}) denotes the logarithmic derivative of the linear growth factor: $f \equiv \frac{\mathrm{d} \ln D_1}{\mathrm{d} \ln a}$. $u(k|M)$ and $u_g(k|M)$ are the normalized Fourier transforms of the dark matter and galaxy density distributions within a halo of mass $M$. In our calculations we take both of these distributions given by the NFW profile \citep{1997ApJ...490..493N} and the concentration parameter $c(M)$ is taken from \citet{2001MNRAS.321..559B}. One dimensional velocity dispersion of the galaxies inside a halo with mass $M$ is taken to follow the scaling of the isothermal sphere model:
\begin{equation}\label{eq_b3}
\sigma = \gamma \sqrt {\frac{GM}{2R_\mathrm{vir}}}\,,
\end{equation}
where $R_\mathrm{vir}$ is the virial radius of the halo and $\gamma$ is a free parameter.  

After specifying the background cosmology the above described model has four free parameters: $M_0$, $\alpha$ (Eq. (\ref{eq_b2})), $\sigma$ (Eq. (\ref{eq_b3})) and $M_\mathrm{low}$. The last parameter $M_\mathrm{low}$ represents the lower boundary of the mass integration i.e. halos with masses below $M_\mathrm{low}$ are assumed to be ``dark''.

One can also use the halo model to estimate nonlinear contributions to the power spectrum covariance matrix. The additional term to the covariance matrix $C_{ij}^{NL}$ ($i,j-$denote power spectrum bins) arising from the parallelogram configurations of the trispectrum \footnote{Here only the contribution due to the 1-halo term is given.} is given by \citep{2004MNRAS.348..250C}: 
\begin{eqnarray}
& & C_{ij}^{NL} = \frac{T_{ij}}{V}=\frac{1}{V}\int \limits_i \frac{\mathrm{d}^3 k\,}{V_i} \int \limits_j \frac{\mathrm{d}^3 k\,}{V_j} \int \mathrm{d}M\,n(M) \cdot\nonumber \\ 
& & \frac{\langle N(N-1)(N-2)(N-3)|M \rangle}{\bar{n}^4} |u_g(k_i|M)|^2 |u_g(k_j|M)|^2\,. \label{eq_b6}
\end{eqnarray}
$\int \limits_i$ denotes an integral over a $k$-space shell centered at wavenumber $k_i$ with a volume $V_i=4\pi k_i^2 \Delta k$. The 4th moment of the halo occupation distribution is taken as:
\begin{eqnarray}
\langle N(N-1)(N-2)(N-3)|M \rangle = \beta^2(M)\left[2\beta^2(M)-1\right]\cdot \nonumber \\
\left[3\beta^2(M)-2\right]\langle N|M \rangle^4\,, 
\end{eqnarray}
where $\beta(M)$ and $\langle N|M \rangle$ are given in Eqs. (\ref{eq_b4}) and (\ref{eq_b2}) above. Performing calculations in redshift space a factor of $\mathcal{R}_p^2(k\sigma)$ (see Eq. (\ref{eq_b5})) must also be included in Eq. (\ref{eq_b6}).

\section{Fitting formulae for the coupling kernels}\label{appd}
In this appendix we provide analytical fitting formulae for the coupling kernels $K(k,k')$ in Eq. (\ref{eq10}) \footnote{To avoid confusion we do not call them window functions since the word ``window'' has been already used to mean the Fourier transform of the survey volume.}. The analytic form is motivated by the fact that the angle averaged survey window $|W(k)|^2$ (see Fig. \ref{fig8}) can be reasonably well approximated by the analytical form:
\begin{equation}\label{eq_d1}
|W(k)|^2 \equiv f(k) = \frac{1}{1 + \left(\frac{k}{a}\right)^2 + \left(\frac{k}{b}\right)^4}\,.
\end{equation}
Now assuming that $|W(\mathbf{k})|^2$ is isotropic (which certainly is not the case as seen from Fig. \ref{fig9}), we can find the coupling kernels $K(k,k')$ as:
\begin{equation}\label{eq_d2}
K(k,k') = C\cdot\int{\rm d}\Omega_\mathbf{k}\int{\rm d}\Omega_\mathbf{k'} |W(\mathbf{k}-\mathbf{k'})|^2=C\cdot\frac{8\pi^2}{kk'}\int \limits_{|k-k'|}^{k+k'}f(x)x{\rm d}x\,. 
\end{equation}
For $f(k)$ given by Eq. (\ref{eq_d1}) the integral in Eq. (\ref{eq_d2}) and the normalization constant $C$ can be found analytically. The kernels are normalized such that
\begin{equation}
\int K(k,k')k'^2{\rm d}k' = 1
\end{equation}
is satisfied.

Depending on the values of $a$ and $b$ there are two different solutions.
\begin{enumerate}
\item $b^4>4a^4$:
\begin{equation}
K(k,k') = \frac{C}{kk'}\ln \left[ \frac{g(k-k')}{g(k+k')} \right]\,,
\end{equation}
where
\begin{eqnarray}
g(x) = \frac{\mu + b^4 + 2a^2x^2}{\mu - b^4 - 2a^2x^2}\,,\\
\mu = b^2\sqrt{b^4 - 4a^4}
\end{eqnarray}
and the normalization constant:
\begin{equation}
C = \frac{a}{\pi\sqrt{2}\left(\sqrt{b^4+\mu}-\sqrt{b^4-\mu}\right)}\,.
\end{equation}
\item $b^4<4a^4$:

\begin{equation}
K(k,k') = \frac{C}{kk'} \left[ g(k+k') - g(k-k') \right]\,,
\end{equation}
where
\begin{equation}
g(x) = \arctan\left(\frac{b^4 + 2a^2x^2}{b^2\sqrt{4a^4-b^4}}\right)
\end{equation}
and the normalization constant:
\begin{equation}
C = \frac{1}{\pi b \sqrt{2-\left(\frac{b}{a}\right)^2}}\,.
\end{equation}
\end{enumerate}     
Although the isotropy assumption is certainly not correct, the above parametric family provides a very good fit to the numerically evaluated kernels as seen in Fig. \ref{fig10}. The best fitting $a$ and $b$ for the analyzed SDSS LRG sample are $0.00457$ and $0.00475$, respectively. 

\section{Nonlinear model fitting. Correlated data}\label{appe}
We find the best fitting parameters for the nonlinear model by minimizing $\chi ^2$, which in the case of Gaussian errors is equivalent to finding the maximum likelihood solution. For this purpose we use Levenberg-Marquardt method as described in \citet{1992nrfa.book.....P}, where it was assumed that data values are uncorrelated. Since we are interested in the case with correlated errors, we have to make slight modifications to their implementation of the algorithm.   

Using their notation, $\chi ^2$ is now calculated as:
\begin{equation}\label{eq_e1}
\chi ^2(\mathbf{a})=\sum \limits_{i=1}^{N} \sum \limits_{j=1}^{N} \left [ {y_i - y(x_i;\mathbf{a})} \right ]\cdot C^{-1}_{ij} \cdot \left [ {y_j - y(x_j;\mathbf{a})} \right ]\,,
\end{equation}
and the quantities $\beta_k$ and $\alpha_{kl}$ as follows:
\begin{equation}
\beta_k=\sum \limits_{i=1}^{N} \sum \limits_{j=1}^{N} \left [ {y_i - y(x_i;\mathbf{a})} \right ]\cdot C^{-1}_{ij} \cdot \frac{\partial y(x_j;\mathbf{a})}{\partial a_k}\,,
\end{equation}
\begin{equation}
\alpha_{kl}=\sum \limits_{i=1}^{N} \sum \limits_{j=1}^{N}
  \frac{\partial y(x_i;\mathbf{a})}{\partial a_k} \cdot C^{-1}_{ij} \cdot \frac{\partial y(x_j;\mathbf{a})}{\partial a_l}\,.
\end{equation}
In the above relations $C_{ij}$ represents the data covariance matrix.

\section{Goodness of fit. Correlated Gaussian data}\label{appf}
Under the assumption that statistical fluctuations $\Delta y_i = y_i - y(x_i;\mathbf{a})$ $(i=1\ldots N)$ in Eq. (\ref{eq_e1}) are Gaussian distributed, with covariance matrix $C_{ij}$, one can easily derive probability density function (pdf) for the quantity $\chi^2$, and thus open up a way to estimate the goodness of fit. $\chi^2$ goodness-of-fit estimator is usually exploited in the case of independent Gaussian variables. Here we show that calculating $\chi^2$ for the correlated Gaussian data as given in Eq. (\ref{eq_e1}), one obtains the same result that is well known for the independently distributed case.

According to our assumption $\mathbf{\Delta y}$ is Gaussian distributed:
\begin{equation}
f_{\mathbf{\Delta y}}(\mathbf{\Delta y})=\frac{1}{\sqrt{(2\pi)^N \det \mathbf{C}}}\exp\left(-\frac{1}{2}\mathbf{\Delta y}^T\cdot \mathbf{C}^{-1} \cdot\mathbf{\Delta y}\right).
\end{equation}
The conditional pdf of $\chi^2$ given $\mathbf{\Delta y}$:
\begin{equation}
f_{\chi^2|\mathbf{\Delta y}}(\chi^2|\mathbf{\Delta y})=\delta\left(\chi^2 - \mathbf{\Delta y}^T\cdot\mathbf{C}^{-1}\cdot\mathbf{\Delta y}\right),
\end{equation}
and so the pdf for $\chi^2$ can be written as:
\begin{eqnarray}
f_{\chi^2}(\chi^2)=\frac{1}{\sqrt{(2\pi)^N \det \mathbf{C}}}\int{\rm d}^N\!\Delta y\, \exp\left(-\frac{1}{2}\mathbf{\Delta y}^T\cdot\mathbf{C}^{-1}\cdot\mathbf{\Delta y}\right)\cdot\nonumber \\ 
\delta\left(\chi^2 - \mathbf{\Delta y}^T\cdot\mathbf{C}^{-1}\cdot\mathbf{\Delta y}\right). 
\end{eqnarray}
Now we define a new set of variables:
\begin{equation}
\mathbf{\Delta y}' = \mathbf{L}^T\cdot\mathbf{\Delta y},
\end{equation}
where $\mathbf{L}$ is the lower triangular matrix appearing in the Cholesky decomposition of $\mathbf{C}^{-1}$:
\begin{equation}
\mathbf{C}^{-1} = \mathbf{L}\cdot\mathbf{L}^T.
\end{equation}
Since $\mathbf{C}^{-1}$ can be seen as the metric tensor, we can write for the transformation of the volume elements \footnote{The metric in the new frame is an identity matrix.}:
\begin{equation}
\sqrt{\det \mathbf{C}^{-1}}{\rm d}^N\!\Delta y = \frac{{\rm d}^N\!\Delta y}{\sqrt{\det \mathbf{C}}}={\rm d}^N\!\Delta y'.
\end{equation}
In the new frame, after changing to the spherical coordinates and integration over the angles:
\begin{equation}
f_{\chi^2}(\chi^2)=(2\pi)^{-\frac{N}{2}}\cdot\frac{\Omega_N}{2}\cdot\int{\rm d}(\Delta y'^2)\,\Delta y'^{N-2}\exp\left(-\frac{\Delta y'^2}{2}\right)\cdot\delta\left(\chi^2 - \Delta y'^2\right),
\end{equation}
where the total $N-$dimensional solid angle:
\begin{equation}
\Omega_N=\frac{2\pi^{\frac{N}{2}}}{\Gamma\left(\frac{N}{2}\right)}.
\end{equation}
Thus the final result reads as:
\begin{equation}
f_{\chi^2}(\chi^2)=\frac{1}{2^{\frac{N}{2}}\Gamma\left(\frac{N}{2}\right)}\left(\chi^2\right)^{\frac{N}{2}-1}\exp\left(-\frac{\chi^2}{2}\right),
\end{equation}
which is a chi-square distribution with $N$ degrees of freedom. Fitting $P$ parameters (equivalent to adding $P$ constraints), the effective number of degrees of freedom drops to $N_{\mathrm{eff}}=N-P$, as usual.

Now it is straightforward to calculate $p-$values describing the goodness of fit.

\section{SDSS LRG power spectrum and covariance matrix}\label{appg}

\begin{deluxetable}{cccccccccccccccccccc}
\tabletypesize{\tiny}
\rotate
\tablecolumns{20}
\tablewidth{0pc}
\tablecaption{Measured SDSS LRG power spectrum and covariance matrix from 1000 mock catalogs.\label{tab1}}
\tablehead{
\colhead{bin \#} &
\colhead{$k_i\,[h\,\mathrm{Mpc}^{-1}]$} &
\colhead{$P\,[h^{-3}\,\mathrm{Mpc}^{3}]$} &
\colhead{$\Delta P\,[h^{-3}\,\mathrm{Mpc}^{3}]$} &
\colhead{1} & 
\colhead{2} & 
\colhead{3} & 
\colhead{4} & 
\colhead{5} & 
\colhead{6} & 
\colhead{7} & 
\colhead{8} & 
\colhead{9} & 
\colhead{10} & 
\colhead{11} & 
\colhead{12} & 
\colhead{13} & 
\colhead{14} & 
\colhead{15} & 
\colhead{16}
}
\startdata
1 & 0.148e-1 & 0.133e6 & 0.228e5 & 0.518e9 & 0.358e8 & 0.762e7 & 0.521e7 & 0.337e7 & 0.161e7 & 0.189e7 & 0.987e6 & 0.128e7 & 0.106e7 & 0.400e6 & 0.246e6 & 0.204e6 & 0.114e6 & 0.106e6 & 0.184e6 \\
2 & 0.315e-1 & 0.712e5 & 0.461e4 & 0.358e8 & 0.212e8 & 0.267e7 & 0.122e7 & 0.433e6 & 0.402e6 & 0.260e6 & 0.184e6 & 0.198e6 & 0.165e6 & 0.146e6 & 0.124e6 & 0.958e5 & 0.854e5 & 0.420e5 & 0.317e5 \\
3 & 0.504e-1 & 0.500e5 & 0.224e4 & 0.762e7 & 0.267e7 & 0.500e7 & 0.103e7 & 0.362e6 & 0.316e6 & 0.192e6 & 0.170e6 & 0.810e5 & 0.702e5 & 0.515e5 & 0.467e5 & 0.389e5 & 0.285e5 & 0.138e5 & 0.151e5 \\
4 & 0.697e-1 & 0.394e5 & 0.140e4 & 0.521e7 & 0.122e7 & 0.103e7 & 0.196e7 & 0.399e6 & 0.204e6 & 0.159e6 & 0.142e6 & 0.117e6 & 0.986e5 & 0.839e5 & 0.614e5 & 0.652e5 & 0.492e5 & 0.152e5 & 0.215e5 \\
5 & 0.891e-1 & 0.286e5 & 0.816e3 & 0.337e7 & 0.433e6 & 0.362e6 & 0.399e6 & 0.667e6 & 0.160e6 & 0.112e6 & 0.110e6 & 0.788e5 & 0.654e5 & 0.537e5 & 0.647e5 & 0.564e5 & 0.477e5 & 0.266e5 & 0.298e5 \\
6 & 0.109e0 & 0.213e5 & 0.550e3 & 0.161e7 & 0.402e6 & 0.316e6 & 0.204e6 & 0.160e6 & 0.303e6 & 0.113e6 & 0.798e5 & 0.614e5 & 0.520e5 & 0.416e5 & 0.427e5 & 0.422e5 & 0.325e5 & 0.233e5 & 0.265e5 \\
7 & 0.128e0 & 0.185e5 & 0.462e3 & 0.189e7 & 0.260e6 & 0.192e6 & 0.159e6 & 0.112e6 & 0.113e6 & 0.213e6 & 0.817e5 & 0.594e5 & 0.436e5 & 0.384e5 & 0.284e5 & 0.299e5 & 0.256e5 & 0.215e5 & 0.217e5 \\
8 & 0.148e0 & 0.147e5 & 0.364e3 & 0.987e6 & 0.184e6 & 0.170e6 & 0.142e6 & 0.110e6 & 0.798e5 & 0.817e5 & 0.133e6 & 0.533e5 & 0.366e5 & 0.312e5 & 0.269e5 & 0.217e5 & 0.188e5 & 0.181e5 & 0.129e5 \\
9 & 0.167e0 & 0.119e5 & 0.295e3 & 0.128e7 & 0.198e6 & 0.810e5 & 0.117e6 & 0.788e5 & 0.614e5 & 0.594e5 & 0.533e5 & 0.871e5 & 0.384e5 & 0.266e5 & 0.242e5 & 0.217e5 & 0.188e5 & 0.128e5 & 0.143e5 \\
10 & 0.187e0 & 0.103e5 & 0.251e3 & 0.106e7 & 0.165e6 & 0.702e5 & 0.986e5 & 0.654e5 & 0.520e5 & 0.436e5 & 0.366e5 & 0.384e5 & 0.631e5 & 0.254e5 & 0.200e5 & 0.169e5 & 0.136e5 & 0.927e4 & 0.100e5 \\
11 & 0.206e0 & 0.869e4 & 0.224e3 & 0.400e6 & 0.146e6 & 0.515e5 & 0.839e5 & 0.537e5 & 0.416e5 & 0.384e5 & 0.312e5 & 0.266e5 & 0.254e5 & 0.504e5 & 0.237e5 & 0.181e5 & 0.158e5 & 0.114e5 & 0.123e5 \\
12 & 0.226e0 & 0.759e4 & 0.210e3 & 0.246e6 & 0.124e6 & 0.467e5 & 0.614e5 & 0.647e5 & 0.427e5 & 0.284e5 & 0.269e5 & 0.242e5 & 0.200e5 & 0.237e5 & 0.441e5 & 0.217e5 & 0.174e5 & 0.148e5 & 0.141e5 \\
13 & 0.246e0 & 0.654e4 & 0.199e3 & 0.204e6 & 0.958e5 & 0.389e5 & 0.652e5 & 0.564e5 & 0.422e5 & 0.299e5 & 0.217e5 & 0.217e5 & 0.169e5 & 0.181e5 & 0.217e5 & 0.396e5 & 0.191e5 & 0.133e5 & 0.146e5 \\
14 & 0.265e0 & 0.599e4 & 0.182e3 & 0.114e6 & 0.854e5 & 0.285e5 & 0.492e5 & 0.477e5 & 0.325e5 & 0.256e5 & 0.188e5 & 0.188e5 & 0.136e5 & 0.158e5 & 0.174e5 & 0.191e5 & 0.330e5 & 0.152e5 & 0.120e5 \\
15 & 0.285e0 & 0.530e4 & 0.162e3 & 0.106e6 & 0.420e5 & 0.138e5 & 0.152e5 & 0.266e5 & 0.233e5 & 0.215e5 & 0.181e5 & 0.128e5 & 0.927e4 & 0.114e5 & 0.148e5 & 0.133e5 & 0.152e5 & 0.261e5 & 0.155e5 \\
16 & 0.305e0 & 0.479e4 & 0.164e3 & 0.184e6 & 0.317e5 & 0.151e5 & 0.215e5 & 0.298e5 & 0.265e5 & 0.217e5 & 0.129e5 & 0.143e5 & 0.100e5 & 0.123e5 & 0.141e5 & 0.146e5 & 0.120e5 & 0.155e5 & 0.271e5 \\
\enddata
\end{deluxetable}

\end{document}